\documentclass[journal]{IEEEtran}
\ifCLASSINFOpdf
\else
\fi
\usepackage{url}

\usepackage{graphics} 
\usepackage{epsfig} 
\usepackage{mathptmx} 
\usepackage{times} 
\usepackage{amsmath} 
\usepackage{amssymb}  
\usepackage[export]{adjustbox}
\usepackage{tikz} 
\usepackage{pgfplots} 
\pgfplotsset{grid style={dotted,black}} 
\pgfplotsset{compat=newest} 

\begin{document}
%
\title{Deep Learning Methods for Improved Decoding of Linear Codes}
%
%
%

\author{Eliya~Nachmani,
	    Elad~Marciano,
	    Loren~Lugosch,~\IEEEmembership{Member,~IEEE,}
	    Warren~J.~Gross,~\IEEEmembership{Senior~Member,~IEEE,}
	    David~Burshtein,~\IEEEmembership{Senior~Member,~IEEE,}
        and~Yair~Be'ery,~\IEEEmembership{Senior~Member,~IEEE} 
\thanks{E. Nachmani, E. Marchiano, D. Burshtein and Y. Be'ery are with the School
of Electrical Engineering, Tel-Aviv University, Tel-Aviv, 6997801 Israel, e-mail: enk100@gmail.com, eladmarc@gmail.com, burstyn@eng.tau.ac.il, ybeery@eng.tau.ac.il.}
\thanks{L. Lugosch and W. J. Gross are with the Department
of Electrical and Computer Engineering, McGill University, Montr\'eal, QC H3A 0G4, Canada, e-mail: loren.lugosch@mail.mcgill.ca, warren.gross@mcgill.ca.}
\thanks{This work was presented in part in the Allerton 2016 conference and in ISIT 2017.}
}

%
%

\markboth{Accepted to IEEE Journal of Selected Topics in Signal Processing}%
{Nachmani \MakeLowercase{\textit{et al.}}: Deep Learning Methods for Improved Decoding of Linear Codes}
%



\maketitle

\begin{abstract}
The problem of low complexity, close to optimal, channel decoding of linear codes with short to moderate block length is considered. It is shown that deep learning methods can be used to improve a standard belief propagation decoder, despite the large example space. Similar improvements are obtained for the min-sum algorithm. It is also shown that tying the parameters of the decoders across iterations, so as to form a recurrent neural network architecture, can be implemented with comparable results. The advantage is that significantly less parameters are required. We also introduce a recurrent neural decoder architecture based on the method of successive relaxation.
Improvements over standard belief propagation are also observed on sparser Tanner graph representations of the codes. Furthermore, we demonstrate that the neural belief propagation decoder can be used to improve the performance, or alternatively reduce the computational complexity, of a close to optimal decoder of short BCH codes.
\end{abstract}

\begin{IEEEkeywords}
Deep learning, error correcting codes, belief propagation, min-sum decoding.
\end{IEEEkeywords}

%
\IEEEpeerreviewmaketitle

\section{Introduction}
In recent years deep learning methods have demonstrated amazing performances in various tasks. 
These methods outperform human-level object detection in some tasks \cite{resnet}, they achieve state-of-the-art results in machine translation \cite{nmt} and speech processing \cite{graves2013speech}, and they attain record breaking performances in challenging games such as Go \cite{d_silver}. 

In this paper we suggest an application of deep learning methods to the problem of low complexity channel decoding.
A well-known family of linear error correcting codes are the linear low-density parity-check (LDPC) codes \cite{galmono}. LDPC codes achieve near Shannon channel capacity with the belief propagation (BP) decoding algorithm, but can typically do so for relatively large block lengths.
For short to moderate high density parity check (HDPC) codes \cite{jiang2006iterative,dimnik2009improved,yufit2011efficient,zhang2012adaptive,helmling2014efficient}, such as common powerful linear algebraic codes, the regular BP algorithm obtains poor results compared to the optimal maximum likelihood (ML) decoder. On the other hand, the importance of close to optimal low complexity, low latency and low power decoders of short to moderate codes has grown with the emergence of applications driven by the Internet of Things.

Recently, in \cite{nachmani} it has been shown that deep learning methods can improve the BP decoding of HDPC codes using a weighted BP decoder. The BP algorithm is formulated as a neural network and it is shown that it can improve the decoding by $0.9{\rm dB}$ in the high SNR regime. A key property of the method is that it is sufficient to train the neural network decoder using a single codeword (e.g., the all-zero codeword), since the architecture guarantees the same error rate for any chosen transmitted codeword.
Later, Lugosch \& Gross \cite{lugosch} proposed an improved neural network architecture that achieves similar results to \cite{nachmani} with less parameters and reduced complexity. The main difference compared to \cite{nachmani} is that the offset min-sum algorithm is used instead of the sum-product algorithm, thus eliminating the need to use multiplications. Gruber et al. \cite{tenbrink} proposed a neural network decoder with an unconstrained graph (i.e., fully connected network) and showed that the network gets close to the ML performance for very small block codes, $N=16$. Also, O'Shea \& Hoydis \cite{AutoencoderComm} proposed to use an autoencoder as a communication system for small block code with $N=7$. In \cite{cammerer2017scaling} it was suggested to improve an iterative decoding algorithm of polar codes by using neural network decoders of sub-blocks. In \cite{farsad2017detection} deep learning-based detection algorithms were used when the channel model is unknown, and in \cite{samuel2017deep} deep learning was used for MIMO detection. Deep learning was also applied to quantum error correcting codes \cite{krastanov2017deep}.

In this work we elaborate on our work in \cite{nachmani} and \cite{lugosch} and extend it as follows\footnote{See the preprint \cite{nachmani2017rnn}.}. First, we apply tying to the decoder parameters by using a recurrent neural network (RNN) architecture, and show that it can achieve up to $1.5{\rm dB}$ improvement over the standard belief propagation algorithm in the high SNR regime. The advantage over the feed-forward architecture in our initial work \cite{nachmani} is that it reduces the number of parameters. Similar improvements were obtained when applying tying to the neural min-sum algorithms. We introduce a new RNN decoder architecture based on the successive relaxation technique and show that it can achieve excellent performance with just a single learnable parameter.
We also investigate the performance of the RNN decoder on parity check matrices with lower densities and fewer short cycles and show that despite the fact that we start with reduced cycle matrix, the network can improve the performance up to $1.0{\rm dB}$. The output of the training algorithm can be interpreted as a soft Tanner graph that replaces the original one. State of the art decoding algorithms of short to moderate algebraic codes, such as \cite{fossOSD1,dimnik2009improved,helmling2014efficient}, utilize the BP algorithm as a component in their solution. Thus, it is natural to replace the standard BP decoder with our trained RNN decoder, in an attempt to improve either the decoding performance or its complexity. In this work we demonstrate, for a BCH(63,36) code, that such improvements can be realized by using RNN decoders in the mRRD algorithm \cite{dimnik2009improved}.

\section{Trellis representation of belief propagation}
The renowned BP decoder \cite{galmono}, \cite{ru_book} can be constructed from the Tanner graph, which is a graphical representation of some parity check matrix that describes the code. In this algorithm, messages are transmitted over edges. Each node calculates its outgoing transmitted message over some edge, based on all incoming messages it receives over all the other edges. We start by providing an alternative graphical representation to the BP algorithm with $L$ full iterations when using parallel (flooding) scheduling. Our alternative representation is a trellis in which the nodes in the hidden layers correspond to edges in the Tanner graph.  
Denote by $N$, the code block length (i.e., the number of variable nodes in the Tanner graph), and by $E$, the number of edges in the Tanner graph. Then the input layer of our trellis representation of the BP decoder is a vector of size $N$, that consists of the log-likelihood ratios (LLRs) of the channel outputs. The LLR value of variable node $v$, $v=1,2,\ldots,N$, is given by
$$
l_v = \log\frac{\Pr\left(C_v=1 | y_v\right)}{\Pr\left(C_v=0 | y_v\right)}
$$
where $y_v$ is the channel output corresponding to the $v$th codebit, $C_v$. 

All the following layers in the trellis, except for the last one (i.e., all the hidden layers), have size $E$. For each hidden layer, each processing element in that layer is associated with the message transmitted over some edge in the Tanner graph. The last (output) layer of the trellis consists of $N$ processing elements that output the final decoded codeword. Consider the $i$th hidden layer, $i=1,2,\ldots,2L$. For odd (even, respectively) values of $i$, each processing element in this layer outputs the message transmitted by the BP decoder over the corresponding edge in the graph, from the associated variable (check) node to the associated check (variable) node. A processing element in the first hidden layer ($i=1$), corresponding to the edge $e=(v,c)$, is connected to a single input node in the input layer: It is the variable node, $v$, associated with that edge.
Now consider the $i$th ($i>1$) hidden layer. For odd (even, respectively) values of $i$, the processing node corresponding to the edge $e=(v,c)$ is connected to all processing elements in layer $i-1$ associated with the edges $e'=(v,c')$ for $c'\ne c$ ($e'=(v',c)$ for $v'\ne v$, respectively). For odd $i$, a processing node in layer $i$, corresponding to the edge $e=(v,c)$, is also connected to the $v$th input node.

The BP messages transmitted over the trellis graph are the following. Consider hidden layer $i$, $i=1,2,\ldots,2L$, and let $e=(v,c)$ be the index of some processing element in that layer. We denote by $x_{i,e}$, the output message of this processing element. For odd (even, respectively), $i$, this is the message produced by the BP algorithm after $\lfloor (i-1)/2 \rfloor$ iterations, from variable to check (check to variable) node.

For odd $i$ and $e=(v,c)$ we have (recall that the self LLR message of $v$ is $l_v$),
\begin{equation}
	x_{i,e=(v,c)} = l_v + \sum_{e'=(v,c'),\: c'\ne c} x_{i-1,e'}
	\label{eq:x_ie_RB}
\end{equation}
under the initialization, $x_{0,e'}=0$ for all edges $e'$ (in the beginning there is no information at the parity check nodes). The summation in~\eqref{eq:x_ie_RB} is over all edges $e'=(v,c')$ with variable node $v$ except for the target edge $e=(v,c)$. Recall that this is a fundamental property of message passing algorithms~\cite{ru_book}.

Similarly, for even $i$ and $e=(v,c)$ we have,
\begin{equation}
	x_{i,e=(v,c)} = 2\tanh^{-1} \left( \prod_{e'=(v',c),\: v'\ne v} \tanh \left( \frac{x_{i-1,e'}}{2} \right) \right)
	\label{eq:x_ie_LB}
\end{equation}

The final $v$th output of the network is given by
\begin{equation}
	o_v = l_v + \sum_{e'=(v,c')} x_{2L,e'}
	\label{eq:ov}
\end{equation}
which is the final marginalization of the BP algorithm.

\section{A neural belief propagation decoder}
We suggest the following parameterized deep neural network decoder that generalizes the BP decoder of the previous section. We use the same trellis representation for the decoder as in the previous section. The difference is that now we assign weights to the edges in the Tanner graph. These weights will be trained using stochastic gradient descent which is the standard method for training neural networks. More precisely, our decoder has the same trellis architecture as the one defined in the previous section. However, Equations~\eqref{eq:x_ie_RB}, \eqref{eq:x_ie_LB} and \eqref{eq:ov} are replaced by
\begin{equation}
	x_{i,e=(v,c)} = \tanh \left(\frac{1}{2}\left(w_{i,v} l_v + \sum_{e'=(v,c'),\: c'\ne c} w_{i,e,e'} x_{i-1,e'}\right)\right)
	\label{eq:x_ie_RB_NN}
\end{equation}
for odd $i$,
\begin{equation}
	x_{i,e=(v,c)} = 2\tanh^{-1} \left( \prod_{e'=(v',c),\: v'\ne v}{x_{i-1,e'}}\right)
	\label{eq:x_ie_LB_NN}
\end{equation}
for even $i$, and
\begin{equation}
	o_v = \sigma \left( w_{2L+1,v} l_v + \sum_{e'=(v,c')} w_{2L+1,v,e'} x_{2L,e'} \right)
	\label{eq:ov_NN}
\end{equation}
where $\sigma(x) \equiv \left( 1+e^{-x} \right)^{-1}$ is a sigmoid function that converts the LLR representation of the message to plain probability. It is easy to verify that the proposed message passing decoding algorithm \eqref{eq:x_ie_RB_NN}-\eqref{eq:ov_NN} satisfies the message passing symmetry conditions \cite[Definition 4.81]{ru_book}. Hence, by \cite[Lemma 4.90]{ru_book}, when transmitting over a binary memoryless symmetric (BMS) channel, the error rate is independent of the transmitted codeword. Therefore, to train the network, it is sufficient to use a database which is constructed by using noisy versions of a single codeword. For convenience we use the zero codeword, which must belong to any linear code. The database reflects various channel output realizations when the zero codeword has been transmitted. The goal is to train the parameters $\left \{ w_{i,v},w_{i,e,e'},w_{i,v,e'} \right \}$ to achieve an $N$ dimensional output word which is as close as possible to the zero codeword.
More precisely, we would like to minimize a cross entropy loss function at the last time step,
\begin{equation}
L{(o,y)}=-\frac{1}{N}\sum_{v=1}^{N}y_{v}\log(o_{v})+(1-y_{v})\log(1-o_{v})
\label{eq:cross_entropy}
\end{equation} 
Here $o_{v}$ and $y_{v}=0$ are the final deep neural network output and the actual $v$th component of the transmitted codeword (which is always the zero codeword during the training).

The network architecture is a non-fully connected neural network. We use stochastic gradient descent to train the parameters.
The motivation behind the new proposed parameterized decoder is that by setting the weights properly, one can compensate for small cycles in the Tanner graph that represents the code. That is, messages sent by parity check nodes to variable nodes can be weighted, such that if a message is less reliable since it is produced by a parity check node with a large number of small cycles in its local neighborhood, then this message will be attenuated properly.

\begin{figure}[thpb]
	\centering
	\includegraphics[width=1\linewidth]{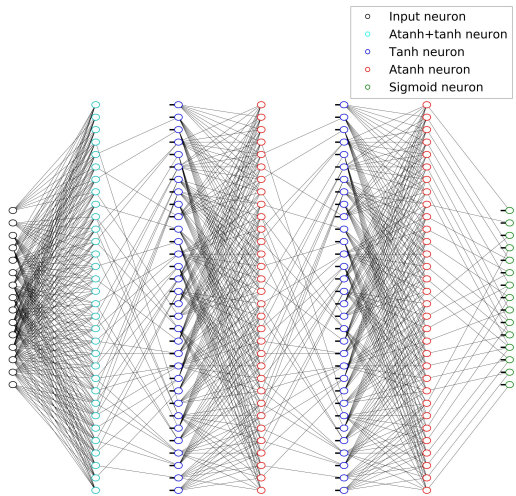}
	\caption{Deep neural network architecture For BCH(15,11) with 5 hidden layers which correspond to 3 full BP iterations. Note that the self LLR messages $l_v$ are plotted as small bold lines. The first hidden layer and the second hidden layer that were described above are merged together. Also note that this figure shows 3 full iterations and the final marginalization.}
	\label{fig:BCH_15_11_arch}
\end{figure}

The time complexity of the deep neural network is roughly the same as the plain BP algorithm, requiring an extra multiplication for each input message. Both have the same number of layers and the same number of non-zero weights in the Tanner graph. The deep neural network architecture is illustrated in Figure~\ref{fig:BCH_15_11_arch} for a BCH(15,11) code.

\section{Neural min-sum decoding}
The standard version of BP described above can be expensive to implement due to the repeated multiplications and hyperbolic functions used to compute the check node function. For this reason, the ``min-sum'' approximation is often used in practical decoder implementations. In min-sum decoding, Equations~\eqref{eq:x_ie_RB} and \eqref{eq:ov} are unchanged, and Equation~\eqref{eq:x_ie_LB} is replaced with Equation~\eqref{eq:x_ie_LB_minsum}:
\begin{equation}
	x_{i,e=(v,c)} = \min_{e'=(v',c),\: v'\ne v}|x_{i-1,e'}| \prod_{e'=(v',c),\: v'\ne v} \text{sign}(x_{i-1,e'})
	\label{eq:x_ie_LB_minsum}
\end{equation}

The min-sum approximation tends to produce messages with large magnitudes, which makes the propagated information seem more reliable than it actually is, causing a BER degradation as a result. To compensate for this effect, the normalized min-sum (NMS) algorithm computes a message using the min-sum approximation, then shrinks the message magnitude using a small weight $w \in (0,1]$, yielding Equation \eqref{eq:x_ie_LB_NMS}:
\begin{equation}
\begin{split}
     x_{i,e=(v,c)} = w \cdot \left( \min_{e'}|x_{i-1,e'}| 
     \prod_{e'} \text{sign}(x_{i-1,e'}) \right), \\e'=(v',c),\: v'\ne v.
\label{eq:x_ie_LB_NMS}
\end{split}
\end{equation}

Similar to the neural BP decoder described above, we propose to assign a learnable weight to each edge and train the decoder as a neural network, yielding a neural normalized min-sum (NNMS) decoder which generalizes the NMS decoder. The check-to-variable messages in NNMS are computed using
\begin{equation}
\begin{split}
     x_{i,e=(v,c)} = w_{i,e=(v,c)} \cdot \left( \min_{e'}|x_{i-1,e'}| 
     \prod_{e'} \text{sign}(x_{i-1,e'}) \right), \\e'=(v',c),\: v'\ne v.
\label{eq:x_ie_LB_NNMS}
\end{split}
\end{equation}
where $w_{i,e=(v,c)}$ is the learnable weight for edge $(v,c)$ in layer $i$. The weights serve a dual purpose: they correct for the min-sum approximation, and, like the weights in the neural BP decoder, they combat the effect of cycles in the Tanner graph.

Both the NNMS decoder and the neural BP decoder require many multiplications, which are generally expensive operations and avoided if possible in a hardware implementation. It was shown in \cite{lugosch} that decoders can learn to improve without using any multiplications at all by adapting the offset min-sum (OMS) algorithm. OMS decoding, like NMS decoding, shrinks a message before sending it, but by subtracting an offset from the message magnitude rather than multiplying by a weight:
\begin{equation}
\begin{split}
     x_{i,e=(v,c)} =  \max\left( \min_{e'}|x_{i-1,e'}| - \beta ,0
      \right) \prod_{e'} \text{sign}(x_{i-1,e'}), \\e'=(v',c),\: v'\ne v.
\label{eq:x_ie_LB_OMS}
\end{split}
\end{equation}
where $\beta$ is the subtracted offset and $\max \left(\dots, 0\right)$ prevents the subtraction from flipping the sign of the message. In the same way that we can generalize NMS to yield NNMS decoding, we can generalize OMS to yield neural offset min-sum (NOMS) decoding, in which check nodes compute the following message:
\begin{equation}
\begin{split}
     x_{i,e=(v,c)} =  \max\left( \min_{e'}|x_{i-1,e'}| - \beta_{i,e=(v,c)},0
      \right) \prod_{e'} \text{sign}(x_{i-1,e'}), \\e'=(v',c),\: v'\ne v.
\label{eq:x_ie_LB_NOMS}
\end{split}
\end{equation}
where $\beta_{i,e=(v,c)}$ is the learnable offset for edge $(v,c)$ in layer $i$.

Note that the functions computed by check nodes in the min-sum decoders are not everywhere differentiable. As a result, the gradient is not defined everywhere. Nevertheless, the functions are non-differentiable only on lower dimensional curves in the space, and are differentiable in the rest of the space. Hence we are applying standard stochastic gradient descent, as is commonly done for neural networks which use activation functions with kinks (see e.g. \cite{lecun1998gradient}), such as rectified linear units (ReLU) which are widely used.

\section{RNN decoding}
\begin{figure}[thpb]
	\centering
	\includegraphics[width=\linewidth]{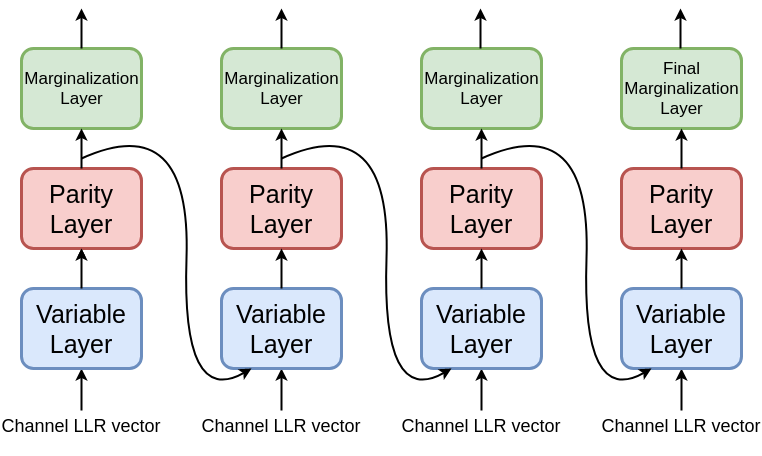}
	\caption{Recurrent Neural Network Architecture with unfold 4 which corresponds to 4 full BP iterations.}
	\label{fig:rnn_unfold}
\end{figure}

We suggest the following parameterized deep neural network decoder which is a constrained version of the BP decoder of the previous section. We use the same trellis representation that was described above for the decoder. The difference is that now the weights of the edges in the Tanner graph are tied, i.e. they are set to be equal in each iteration. This tying transfers the feed-forward architecture that was described above into a recurrent neural network architecture which we term BP-RNN. More precisely, the equations of the proposed architecture are
\begin{align}
\lefteqn{x_{t,e=(v,c)} =} \nonumber \\ 
	&&\tanh \Biggl( \frac{1}{2}  \Biggr. \Biggl(w_{v} l_v + \sum_{e'=(c',v),\: c'\ne c} w_{e,e'} x_{t-1,e'}\Biggr) \Biggl. \Biggr) 
	\label{eq:x_ie_RB_NN_rnn}
\end{align}
where $t=1,2,\ldots$ is the iteration number, $x_{t,e=(v,c)}$ ($x_{t,e=(c,v)}$, respectively) denotes the BP-RNN message from variable node $v$ (check node $c$) to check node $c$ (variable node $v$) at iteration $t$, and
\begin{equation}
	x_{t,e=(c,v)} = 2\tanh^{-1} \left( \prod_{e'=(v',c),\: v'\ne v}{x_{t,e'}}\right)
	\label{eq:x_ie_LB_NN_rnn}
\end{equation}
For iteration $t$ we also have
\begin{equation}
	o_{v,t} = \sigma \left( \tilde{w}_{v} l_v + \sum_{e'=(c',v)} \tilde{w}_{v,e'} x_{t,e'} \right)
	\label{eq:ov_NN_rnn}
\end{equation}
We initialize the algorithm by setting $x_{0,e}=0$ for all $e=(c,v)$. The proposed architecture also preserves the symmetry conditions. As a result the network can be trained by using noisy versions of a single codeword. The training is done as before with a cross entropy loss function, defined in \eqref{eq:cross_entropy}, at the last time step.
The proposed recurrent neural network architecture has the property that after every time step we can add final marginalization and compute the loss of these terms using cross entropy. Using multiloss terms can increase the gradient update of the backpropagation through time algorithm and allow learning the earliest layers. Hence, we suggest the following multiloss variant of \eqref{eq:cross_entropy}:
\begin{equation}
	L{(o,y)}=-\frac{1}{N}\sum_{t=1}^{T}\sum_{v=1}^{N}y_{v}\log(o_{v,t})+(1-y_{v})\log(1-o_{v,t})
	\label{eq:multiloss_cross_entropy}
\end{equation}
where $o_{v,t}$ and $y_{v}=0$ are the deep neural network output at the time step $t$ and the actual $v$th component of the transmitted codeword. This network architecture is illustrated in Figure~\ref{fig:rnn_unfold}. Nodes in the variable layer implement~\eqref{eq:x_ie_RB_NN_rnn}, while nodes in the parity layer implement~\eqref{eq:x_ie_LB_NN_rnn}. Nodes in the marginalization layer implement~\eqref{eq:ov_NN_rnn}. The training goal is to minimize~\eqref{eq:multiloss_cross_entropy}.

Similarly, the neural min-sum decoders can be constrained to use the same weights or offsets for each time step, in which case we drop the iteration index from the learnable parameters, so that Equation~\eqref{eq:x_ie_LB_NNMS} becomes:
\begin{equation}
\begin{split}
     x_{i,e=(v,c)} = w_{e=(v,c)} \cdot \left( \min_{e'}|x_{i-1,e'}| 
     \prod_{e'} \text{sign}(x_{i-1,e'}) \right), \\e'=(v',c),\: v'\ne v,
\label{eq:x_ie_LB_NNMS_RNN}
\end{split}
\end{equation}
and Equation ~\eqref{eq:x_ie_LB_NOMS} becomes:
\begin{equation}
\begin{split}
     x_{i,e=(v,c)} =  \max\left( \min_{e'}|x_{i-1,e'}| - \beta_{e=(v,c)},0
      \right) \prod_{e'} \text{sign}(x_{i-1,e'}), \\e'=(v',c),\: v'\ne v.
\label{eq:x_ie_LB_NOMS_RNN}
\end{split}
\end{equation}

\section{Learning to relax}

Another technique which can be used to improve the performance of belief propagation is the method of successive relaxation (or simply ``relaxation''), as described in \cite{hemati2006dynamics}. In relaxation, an exponentially weighted moving average is applied to combine the message sent in iteration $t-1$ with the raw message computed in iteration $t$ to yield a filtered message, $m_t'$:
\begin{equation}\label{eq:relax}
    m_t' = \gamma m_{t-1}' + (1-\gamma) m_t
\end{equation}
where $\gamma$ is the relaxation factor. As $\gamma \rightarrow 0$, the decoder becomes less relaxed, and as $\gamma \rightarrow 1$, the decoder becomes more relaxed. When $\gamma = 0$, the decoder reverts to being a normal, non-relaxed decoder. 

As it is difficult to predict the behaviour of relaxed decoders analytically, the relaxation factor is chosen through trial-and-error, that is, by simulating the decoder with several possible values of $\gamma$ and choosing the value which leads to the best performance. Rather than choosing the relaxation parameter through brute force simulation, we propose learning this parameter using stochastic gradient descent, as the relaxation operation is differentiable with respect to $\gamma$. Moreover, it is possible to use a separate relaxation parameter for each edge of the Tanner graph, similar to the other decoder architectures described in this work, although we have found that using per-edge relaxation parameters does not improve much upon using a single parameter. To the best of the authors' knowledge, our deep learning methodology constitutes the first technique for optimizing decoder relaxation factors which does not rely on simple trial-and-error.

\begin{figure}
	\centering
	\includegraphics[width=\linewidth]{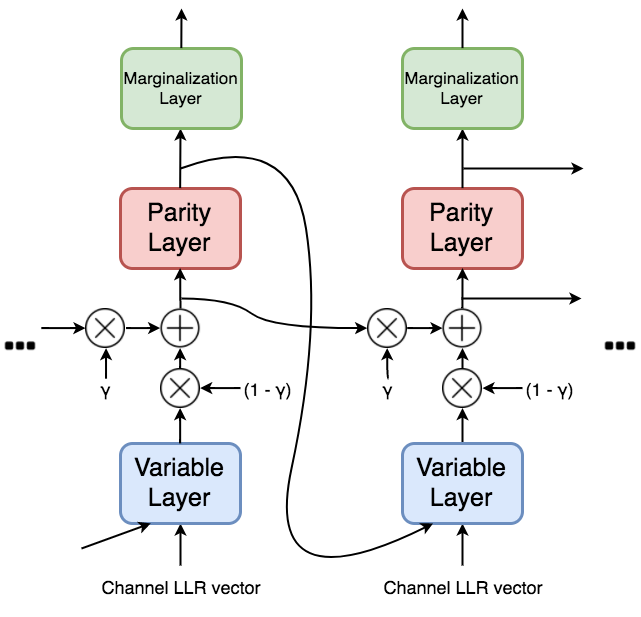}
	\caption{Two iterations of a relaxed decoder.}
	\label{fig:rnn_relaxed}
\end{figure}

The structure of one possible version of a neural decoder with relaxation is illustrated in Figure~\ref{fig:rnn_relaxed}. In this figure, $\times$ indicates elementwise multiplication, $+$ indicates elementwise addition, and $\gamma$ indicates the learnable relaxation parameter(s). Like the decoder shown in Figure~\ref{fig:rnn_unfold}, the relaxed decoder is effectively an RNN, with an additional ``shortcut connection'' (as is found in architectures such as those described in \cite{srivastava2015highway} and \cite{resnet}). Since the relaxation operation can be considered an IIR filter, we require that $\gamma$ be in the range [0,1), as values outside of this range would result in filters with instability or ringing in the impulse response. During training, we use the sigmoid function to squash $\gamma$ into the correct range; during inference, the squashed values can be stored in the decoder so that the sigmoid need not be computed. 

Relaxation is relatively expensive compared to the other neural decoder techniques, as it requires not only multiplications and additions but also additional memory to store the previous iteration's messages. However, as was shown in \cite{relaxed_min_sum}, it is sometimes possible to set relaxation factors to a power of two so that a multiplier-free hardware implementation results, in which case the only additional overhead is memory and additions.

\section{An mRRD algorithm with a neural BP decoder}
\begin{figure}[thpb]
	\centering  \includegraphics[width=0.85\linewidth]{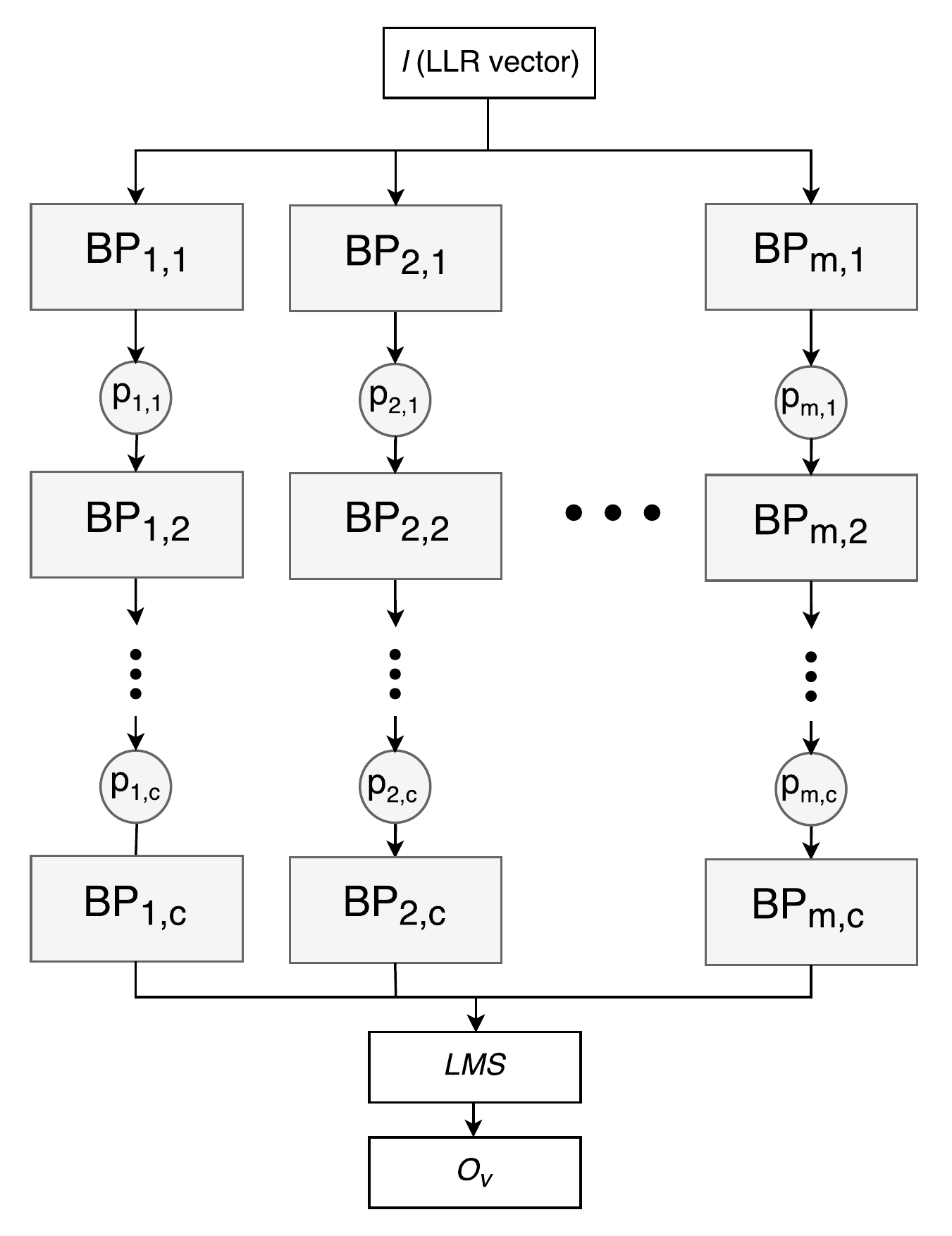}
	\caption{mRRD decoder with $m$ parallel iterative decoders, and $c$ BP blocks in each iterative decoder. The circles represent random permutations from the automorphism group of the code.}
	\label{fig:mrrd_diag}
\end{figure}

Dimnik and Be'ery \cite{dimnik2009improved} proposed a modified random redundant iterative algorithm (mRRD) for decoding HDPC codes based on the RRD \cite{cycle_reduce} and the MBBP \cite{mbbp} algorithms. The mRRD algorithm is a close to ML low complexity decoder for short length ($N<100$) algebraic codes such as BCH codes. This algorithm uses $m$ parallel decoder branches, each comprising of $c$ applications of several (e.g. 2) BP decoding iterations, followed by a random permutation from the automorphism group of the code, as shown in Figure~\ref{fig:mrrd_diag}. The decoding process in each branch stops if the decoded word is a valid codeword. The final decoded word is selected with a least metric selector (LMS) as the one for which the channel output has the highest likelihood. More details can be found in \cite{dimnik2009improved}.

We propose to combine the BP-RNN decoding algorithm with the mRRD algorithm. We can replace the BP blocks in the mRRD algorithm with our BP-RNN decoding scheme. The proposed mRRD-RNN decoder algorithm should achieve near ML performance with less computational complexity.

\section{Experiments And Results} 
The training of all neural networks was performed using TensorFlow~\cite{abadi2016tensorflow}.

\subsection{Neural BP decoders}
We first present results for the neural versions of the standard BP decoder. We apply our method to different linear codes, BCH(63,45), BCH(63,36), BCH(127,64) and BCH(127,99). In all experiments the results on the training, validation and test sets were identical. That is, we did not observe overfitting in our experiments. Details about our experiments
and results are as follows. It should be noted that the parameters $w_{i,v}$ in \eqref{eq:x_ie_RB_NN} and \eqref{eq:ov_NN}, $w_v$ in \eqref{eq:x_ie_RB_NN_rnn} and $\tilde{w}_v$ in \eqref{eq:ov_NN_rnn} were all set to $1$, since training these parameter did not yield additional improvements. Note that these weights multiply the self message from the channel, which is a reliable message, unlike the messages received from check nodes, which may be unreliable due to the presence of short cycles in the Tanner graph. Hence, this self message from the channel can be taken as is.

Training was conducted using stochastic gradient descent with mini-batches. The training data is created by transmitting the zero codeword through a binary input additive white Gaussian noise channel (BIAWGNC) with varying SNRs ranging from $1{\rm dB}$ to $8{\rm dB}$.
We applied the RMSPROP~\cite{rmsprop} rule, using its implementation in~\cite{abadi2016tensorflow}, with learning rate and mini-batch size that depended on the code used and on its parity check matrix. For all BCH codes with $N=63$ the learning rate was $0.001$ and the mini-batch size was $120$. For a BCH(127,99) with right regular parity check matrix, the learning rate was $0.0003$ and the mini-batch size was $80$. For a BCH(127,99) with cycle reduced parity check matrix and for BCH(127,64) the learning rate was $0.003$ and the mini-batch size was $40$. All other parameters were set to their default values in the Tensorflow implementation of the RMSPROP optimizer. As is well known, the training of neural networks requires extensive trial and error tuning of hyper parameters \cite[Appendix A]{srivastava2014dropout}. This was also the case in our experiments that required proper selection of learning rates and mini-batch sizes, taking into account memory limitations of the GPU card as well.
All neural networks had $2$ hidden layers at each time step, and unfold equal to $5$ which corresponds to $5$ full iterations of the BP algorithm. 
At test time, we inject noisy codewords after transmitting through a BIAWGNC and measure the bit error rate (BER) in the decoded codeword at the network output.
The input $x_{t-1,e’}$ to ~\eqref{eq:x_ie_RB_NN_rnn} is clipped such that the absolute value of the input is always smaller than some positive constant $A < 10$. Similar clipping is typically applied in practical implementations of the BP algorithm.\\

\subsubsection{BER For BCH With $N=63$} 

\begin{figure}
	\centering	
	\begin{tikzpicture}
        \begin{semilogyaxis}[
            height=10cm,
            width=9cm,
            ymax=1,
            grid=both,
            xlabel=$E_b / N_0$ (dB),
            ylabel=BER (Bit Error Rate)
        ]

        \addplot coordinates {
            (1, 0.11337599 )
            (2,  0.08164087)
            (3,   0.04941171 )
            (4,  0.02385119 )
            (5,  0.00994147 )
            (6,  0.00348879 )
            (7, 0.00097093  )
            (8, 0.00024583)
        };
        \addlegendentry{BP}

        \addplot coordinates {
            (1, 1.13472222e-01  )
            (2,  8.12843915e-02 )
            (3,   4.58253968e-02  )
            (4,  1.89325397e-02 )
            (5, 5.88095238e-03  )
            (6,  1.32923280e-03 )
            (7,   2.28571429e-04  )
            (8,  2.98941799e-05)
        };
        \addlegendentry{BP-FF}
        
        \addplot coordinates {
            (1, 1.13309524e-01  )
            (2,  8.12817460e-02 )
            (3,   4.66256614e-02 )
            (4,   1.92923280e-02 )
            (5, 5.38624339e-03    )
            (6, 1.21362434e-03   )
            (7, 1.83333333e-04   )
            (8, 1.77248677e-05)
        };
        \addlegendentry{BP-RNN}
        
        \addplot coordinates {
            (1, 1.13958333e-01 )
            (2,  8.26855159e-02  )
            (3,   4.80625000e-02  )
            (4,  1.89117063e-02 )
            (5, 4.87500000e-03  )
            (6,  8.39285714e-04 )
            (7,   9.06746032e-05 )
            (8,   6.54761905e-06)
        };
        \addlegendentry{BP-RNN (Multiloss)}
        
        \addplot coordinates {
            (1, 1.13834656e-01 )
            (2,   8.24378307e-02 )
            (3,   4.66917989e-02 )
            (4,   1.89285714e-02 )
            (5, 5.00529101e-03   )
            (6, 8.21957672e-04   )
            (7, 7.65873016e-05   )
            (8, 5.15873016e-06)
        };
        \addlegendentry{BP-FF (Multiloss)}

        \end{semilogyaxis}
    \end{tikzpicture}
	\caption{BER results for BCH(63,36) code trained with right-regular parity check matrix.}
	\label{fig:bch_63_36_ber_regular}
\end{figure}
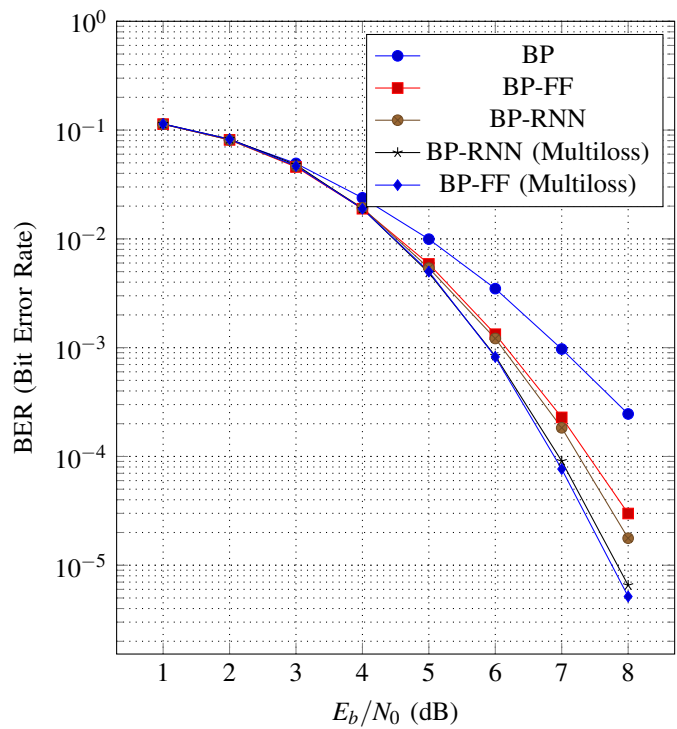

\hfill \break \newline In Figures~\ref{fig:bch_63_45_ber_regular},~\ref{fig:bch_63_36_ber_regular}, we provide the bit-error-rate for a BCH code with $N=63$ using a right-regular parity check matrix based on \cite{parity_g}. As can be seen from the figures, the BP-RNN decoder outperforms the BP feed-forward (BP-FF) decoder by $0.2{\rm dB}$. Not only that we improve the BER, the network has less parameters. Moreover, we can see that the BP-RNN decoder obtains comparable results to the BP-FF decoder when training with multiloss. Furthermore, for the BCH(63,45) and BCH(63,36) there is an improvement up to $1.3{\rm dB}$ and $1.5{\rm dB}$, respectively, over the plain BP algorithm.

\begin{figure}
    \centering
\begin{tikzpicture}
        \begin{semilogyaxis}[
            height=10cm,
            width=9cm,
            ymax=1,
            grid=both,
            xlabel=$E_b / N_0$ (dB),
            ylabel=BER (Bit Error Rate)
        ]
        
        \addplot coordinates {
            (1, 8.88005952e-02)
            (2,   6.10585317e-02 )
            (3,    3.54494048e-02   )
            (4, 1.69206349e-02  )
            (5,    7.01190476e-03 )
            (6,   2.32946429e-03   )
            (7, 6.13591270e-04   )
            (8, 9.51388889e-05)
        };
        \addlegendentry{BP}

        \addplot coordinates {
            (1, 8.83591270e-02 )
            (2,   6.13005952e-02  )
            (3,  3.33591270e-02   )
            (4, 1.32450397e-02 )
            (5,    3.84424603e-03  )
            (6,  8.47023810e-04   )
            (7, 1.36309524e-04   )
            (8, 2.03373016e-05)
        };
        \addlegendentry{BP-FF}
        
        \addplot coordinates {
            (1, 8.90664683e-02 )
            (2,   6.17251984e-02 )
            (3,   3.34394841e-02  )
            (4,  1.29990079e-02 )
            (5,    3.52480159e-03   )
            (6, 7.30952381e-04   )
            (7, 1.05158730e-04   )
            (8, 1.16071429e-05)
        };
        \addlegendentry{BP-RNN}
        
        \addplot coordinates {
            (1,8.96865079e-02   )
            (2,6.26349206e-02   )
            (3,3.49742063e-02   )
            (4,1.32480159e-02)
            (5, 3.16369048e-03   )
            (6,4.91369048e-04   )
            (7,4.88095238e-05   )
            (8,3.96825397e-06)
        };
        \addlegendentry{BP-RNN (Multiloss)}
        
        \addplot coordinates {
            (1, 8.95687831e-02)
            (2,   6.18478836e-02)
            (3,   3.37261905e-02)
            (4,   1.26177249e-02)
            (5,3.08994709e-03   )
            (6,4.67328042e-04   )
            (7,4.77513228e-05  )
            (8, 2.91005291e-06)
        };
        \addlegendentry{BP-FF (Multiloss)}

        \end{semilogyaxis}
    \end{tikzpicture}
    \caption{BER results for BCH(63,45) code trained with right-regular parity check matrix.}
    \label{fig:bch_63_45_ber_regular}
\end{figure}
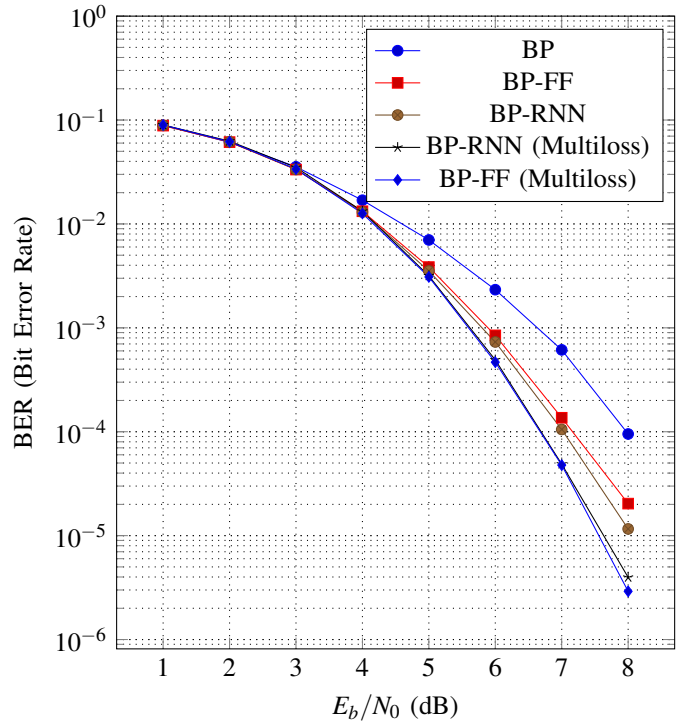

\begin{figure}
	\centering
		\begin{tikzpicture}
        \begin{semilogyaxis}[
            height=10cm,
            width=9cm,
            ymax=1,
            grid=both,
            xlabel=$E_b / N_0$ (dB),
            ylabel=BER (Bit Error Rate)
        ]
        
        \addplot coordinates {
            (1, 8.47390873e-02 )
            (2,   5.25962302e-02 )
            (3,   2.50198413e-02 )
            (4,   7.96726190e-03 )
            (5, 1.60714286e-03   )
            (6, 2.62797619e-04   )
            (7, 2.58928571e-05   )
            (8, 2.28174603e-06)
        };
        \addlegendentry{BP}

        \addplot coordinates {
            (1, 8.53637566e-02 )
            (2,   5.24444444e-02 )
            (3,   2.23240741e-02  )
            (4,  6.86111111e-03 )
            (5, 1.22222222e-03  )
            (6,  1.40873016e-04 )
            (7,   9.39153439e-06 )
            (8,   7.93650794e-07)
        };
        \addlegendentry{BP-FF}
        
        \addplot coordinates {
            (1,8.53703704e-02 )
            (2,   5.29788360e-02 )
            (3,   2.29933862e-02 )
            (4,   6.69708995e-03 )
            (5, 1.26058201e-03  )
            (6,  1.36375661e-04 )
            (7,   8.46560847e-06 )
            (8,   3.96825397e-07)
        };
        \addlegendentry{BP-RNN}
        
        \addplot coordinates {
            (1, 8.56604167e-02 )
            (2,   5.35456349e-02 )
            (3,   2.36110119e-02 )
            (4,   6.42876984e-03 )
            (5,  1.07589286e-03  )
            (6,  1.02380952e-04  )
            (7,  7.53968254e-06  )
            (8,  3.96825397e-07)
        };
        \addlegendentry{BP-RNN (Multiloss)}
        
        \addplot coordinates {
            (1, 8.56349206e-02 )
            (2,   5.28915344e-02 )
            (3,   2.25079365e-02 )
            (4,   6.58862434e-03 )
            (5, 1.03835979e-03 )
            (6,   1.07010582e-04 )
            (7,   7.40740741e-06 )
            (8,   1.32275132e-07)
        };
        \addlegendentry{BP-FF (Multiloss)}
        
        \end{semilogyaxis}
    \end{tikzpicture}
	\caption{BER results for BCH(63,45) code trained with cycle reduced parity check matrix.}
	\label{fig:bch_63_45_ber_iregular}
\end{figure}
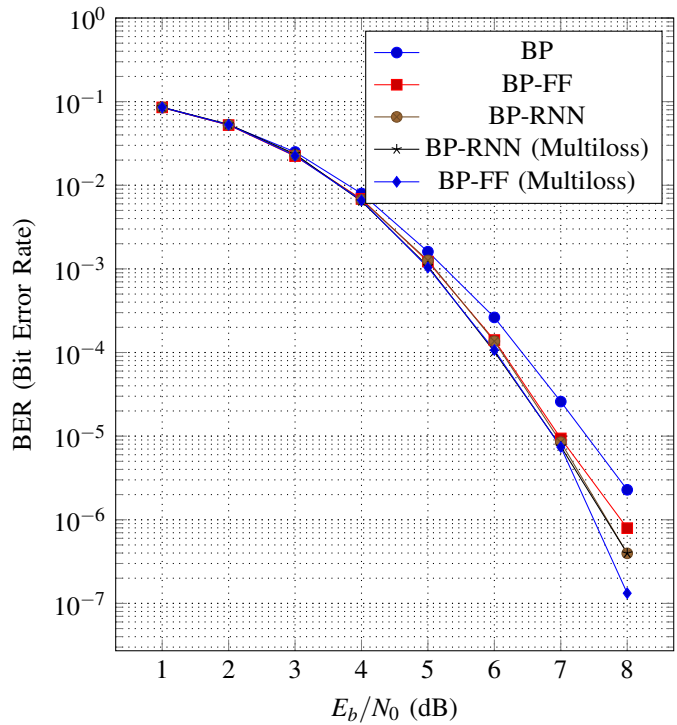

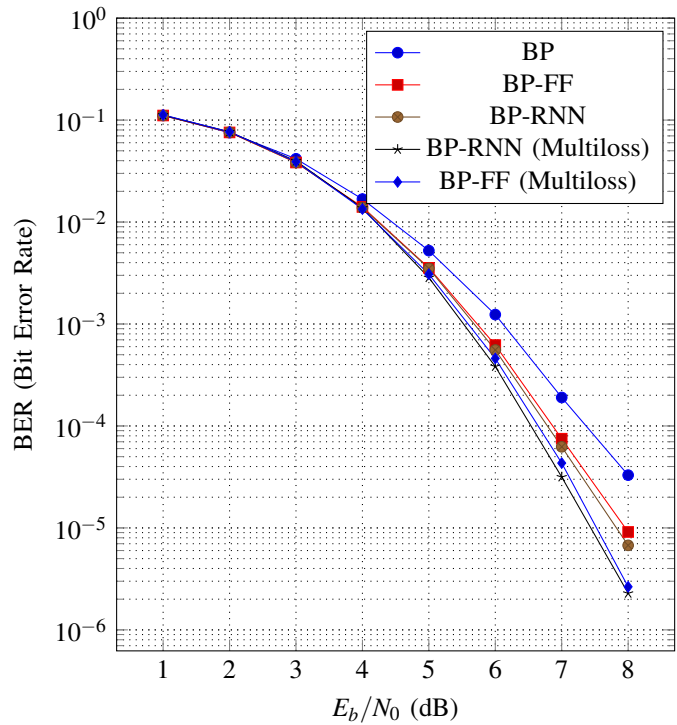
\begin{figure}
	\centering
			\begin{tikzpicture}
        \begin{semilogyaxis}[
            height=10cm,
            width=9cm,
            ymax=1,
            grid=both,
            xlabel=$E_b / N_0$ (dB),
            ylabel=BER (Bit Error Rate)
        ]

        \addplot coordinates {
            (1, 1.11085317e-01 )
            (2,   7.51984127e-02 )
            (3,   4.17242063e-02  )
            (4,  1.68045635e-02 )
            (5,  5.24305556e-03 )
            (6,   1.23680556e-03 )
            (7,   1.90277778e-04 )
            (8,   3.29365079e-05)
        };
        \addlegendentry{BP}

        \addplot coordinates {
            (1, 1.10891534e-01  )
            (2,  7.56878307e-02 )
            (3,   3.83664021e-02 )
            (4,   1.40714286e-02 )
            (5,    3.55687831e-03 )
            (6,   6.21560847e-04  )
            (7,  7.50000000e-05  )
            (8,  9.12698413e-06)
        };
        \addlegendentry{BP-FF}
        
        \addplot coordinates {
            (1, 1.11050265e-01 )
            (2,   7.61230159e-02 )
            (3,   3.84616402e-02 )
            (4,   1.38280423e-02 )
            (5,    3.50925926e-03 )
            (6,   5.56216931e-04 )
            (7,   6.24338624e-05 )
            (8,   6.74603175e-06)
        };
        \addlegendentry{BP-RNN}
        
        \addplot coordinates {
        	(1, 1.11813492e-01 )
        	(2,   7.66220238e-02 )
        	(3,   3.96855159e-02 )
        	(4,   1.36914683e-02 )
        	(5,    2.84722222e-03 )
        	(6,   3.83234127e-04  )
        	(7,  3.15476190e-05  )
        	(8,  2.28174603e-06)
        };
        \addlegendentry{BP-RNN (Multiloss)}
        
        \addplot coordinates {
            (1, 1.12488095e-01 )
            (2,   7.66349206e-02 )
            (3,   3.90859788e-02 )
            (4,   1.34179894e-02)
            (5,    3.08068783e-03 )
            (6,   4.59523810e-04  )
            (7,  4.32539683e-05  )
            (8,  2.64550265e-06)
        };
        \addlegendentry{BP-FF (Multiloss)}
                
        \end{semilogyaxis}
    \end{tikzpicture}
	\caption{BER results for BCH(63,36) code trained with cycle reduced parity check matrix.}
	\label{fig:bch_63_36_ber_iregular}
\end{figure}

In Figures~\ref{fig:bch_63_45_ber_iregular} and~\ref{fig:bch_63_36_ber_iregular}, we provide the bit-error-rate for a BCH code with $N=63$ for a cycle reduced parity check matrix \cite{cycle_reduce}. For BCH(63,45) and BCH(63,36) we get an improvement up to $0.6{\rm dB}$ and $1.0{\rm dB}$, respectively. This observation shows that the neural BP decoder is capable to improve the performance of standard BP even for reduced cycle parity check matrices. Thus answering in the affirmative the uncertainty in \cite{nachmani} regarding the performance of the neural decoder on a cycle reduced parity check matrix. The importance of this finding is that it enables a further improvement in the decoding performance, as BP (both standard BP and the new parameterized BP algorithm) yields lower error rate for sparser parity check matrices. However, as expected, the performance gain of the neural decoder compared to plain BP is lower for a sparser parity check matrix.\\

\subsubsection{BER For BCH With $N=127$}
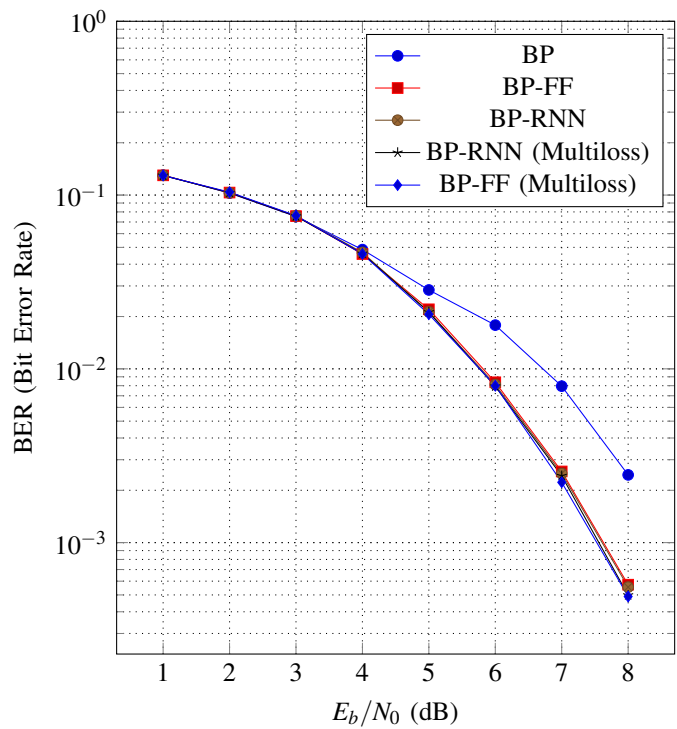
\begin{figure}
	\centering	
	\begin{tikzpicture}
        \begin{semilogyaxis}[
            height=10cm,
            width=9cm,
            ymax=1,
            grid=both,
            xlabel=$E_b / N_0$ (dB),
            ylabel=BER (Bit Error Rate)
        ]

        \addplot coordinates {
            (1, 0.13027165 )
            (2,  0.10261614 )
            (3,  0.07526181 )
            (4,  0.04853937 )
            (5,  0.02844094 )
            (6,  0.01785157 )
            (7,   0.00794902 )
            (8,  0.00245315)
        };
        \addlegendentry{BP}

        \addplot coordinates {
            (1, 0.12989764 )
            (2,  0.10337795 )
            (3,  0.07569291 )
            (4,  0.0457874 )
            (5,   0.02201575 )
            (6,  0.00837185 )
            (7,   0.00256772 )
            (8,  0.00057264)
        };
        \addlegendentry{BP-FF}
        
        \addplot coordinates {
            (1, 0.12990551 )
            (2,  0.1030689 )
            (3,   0.07601575 )
            (4,  0.04671654 )
            (5,  0.02128346 )
            (6,  0.00810748 )
            (7,   0.00249803 )
            (8,  0.00055472)
        };
        \addlegendentry{BP-RNN}
        
        \addplot coordinates {
            (1, 0.12953543 )
            (2,  0.10314567 )
            (3,  0.07548228 )
            (4,  0.04610236 )
            (5,  0.02119882 )
            (6,  0.00792677 )
            (7,   0.00242323 )
            (8,  0.00049803)
        };
        \addlegendentry{BP-RNN (Multiloss)}
        
        \addplot coordinates {
            (1, 0.1296063  )
            (2,  0.10395866 )
            (3,  0.07616929 )
            (4,  0.04581299 )
            (5,  0.0206122 )
            (6,   0.00799213 )
            (7,   0.00222382 )
            (8,  0.00048957)
        };
        \addlegendentry{BP-FF (Multiloss)}
        
        \end{semilogyaxis}
    \end{tikzpicture}
	\caption{BER results for BCH(127,64) code trained with right-regular parity check matrix.}
	\label{fig:bch_127_64_ber_regular}
\end{figure}

\hfill \break \newline In Figure~\ref{fig:bch_127_64_ber_regular}, we provide the bit-error-rate for BCH code with $N=127$ for right-regular parity check matrix based on \cite{parity_g}. As can be seen from the figure, for a right-regular parity check matrix, the BP-RNN and BP-FF decoders obtains an improvement of up to $1.0{\rm dB}$ over the BP, but the BP-RNN decoder uses significantly less parameters compared to BP-FF.

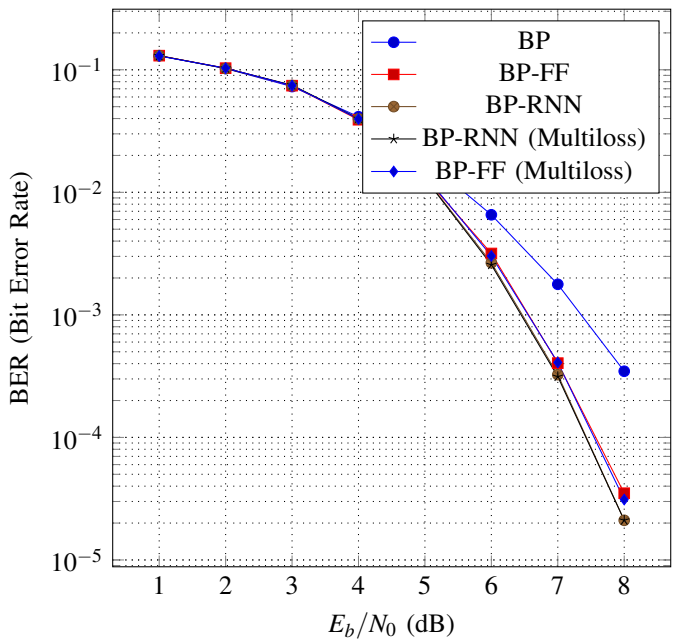
\begin{figure}
	\centering	
	\begin{tikzpicture}
	\begin{semilogyaxis}[
	height=9cm,
	width=9cm,
	grid=both,
	xlabel=$E_b / N_0$ (dB),
	ylabel=BER (Bit Error Rate)
	]

    \addplot coordinates {
		(1, 0.13023622 )
		(2,  0.10232283 )
		(3,  0.07292913 )
		(4,  0.04138386 )
		(5,  0.01860433 )
		(6,  0.0065561 )
		(7,   0.00177579 )
		(8,  0.00034665)
	};
	\addlegendentry{BP}
	
	\addplot coordinates {
		(1, 1.30370079e-01)   
		(2, 1.03064961e-01)   
		(3, 7.41830709e-02)   
		(4, 3.91240157e-02)
		(5, 1.32342520e-02)   
		(6, 3.16889764e-03)   
		(7, 4.03740157e-04)   
		(8, 3.50393701e-05)		
	};
	\addlegendentry{BP-FF}
	
	\addplot coordinates {		
		(1, 1.30358268e-01)   
		(2, 1.02482283e-01)   
		(3, 7.44133858e-02)   
		(4, 3.97440945e-02)
		(5, 1.31791339e-02)   
		(6, 2.66181102e-03)   
		(7, 3.25000000e-04)   
		(8, 2.10629921e-05)
	};
	\addlegendentry{BP-RNN}
		
	\addplot coordinates {
		(1, 1.30100394e-01 )
		(2,   1.02824803e-01 )
		(3,   7.48877953e-02 )
		(4,   4.00452756e-02 )
		(5,   1.30570866e-02 )
		(6,   2.56299213e-03  )
		(7,  3.13188976e-04  )
		(8,  2.10629921e-05)
	};
	\addlegendentry{BP-RNN (Multiloss)}
		
	\addplot coordinates {
		(1, 1.29537402e-01)
		(2, 1.03122047e-01)
		(3, 7.41909449e-02)
		(4, 3.96909449e-02)
		(5, 1.37185039e-02)
		(6, 3.02539370e-03)
		(7, 4.08464567e-04)
		(8, 3.11023622e-05)
	};
	\addlegendentry{BP-FF (Multiloss)}
	
	\end{semilogyaxis}
	\end{tikzpicture}
	\caption{BER results for BCH(127,64) code trained with cycle reduced parity check matrix.}
	\label{fig:bch_127_64_ber_iregular}
\end{figure}

\begin{figure}
	\centering	
	\begin{tikzpicture}
	\begin{semilogyaxis}[
	height=9cm,
	width=9cm,
	grid=both,
	xlabel=$E_b / N_0$ (dB),
	ylabel=BER (Bit Error Rate)
	]
	
	\addplot coordinates {
		(1, 8.06351706e-02)   
		(2, 5.77801837e-02)   
		(3, 3.61824147e-02)   
		(4, 1.77854331e-02)
		(5, 7.58661417e-03)   
		(6, 2.39881890e-03)   
		(7, 5.12664042e-04)   
		(8, 8.73359580e-05)		
	};
	\addlegendentry{BP}
	
	\addplot coordinates {
		(1, 8.08090551e-02)   
		(2, 5.77460630e-02)   
		(3, 3.64291339e-02)   
		(4, 1.57342520e-02)
		(5, 3.99803150e-03)   
		(6, 5.78937008e-04)   
		(7, 5.05905512e-05)   
		(8, 2.75590551e-06)
	};
	\addlegendentry{BP-FF}

	\addplot coordinates {
		(1, 8.09763780e-02)   
		(2, 5.72578740e-02)   
		(3, 3.69212598e-02)   
		(4, 1.64015748e-02)
		(5, 4.35433071e-03)   
		(6, 5.34251969e-04)   
		(7, 4.13385827e-05)   
		(8, 2.16535433e-06)
	};
	\addlegendentry{BP-RNN}
	
	\addplot coordinates {
		(1, 8.16023622e-02)   
		(2, 5.77618110e-02)   
		(3, 3.73011811e-02)   
		(4, 1.63779528e-02)
		(5, 3.81102362e-03)   
		(6, 5.86811024e-04)   
		(7, 3.32677165e-05)   
		(8, 2.55905512e-06)
	};
	\addlegendentry{BP-RNN (Multiloss)}
	
	\addplot coordinates {
		(1, 8.10649606e-02)   
		(2, 5.80078740e-02)   
		(3, 3.66653543e-02)   
		(4, 1.57893701e-02)
		(5, 4.01377953e-03)   
		(6, 5.93897638e-04)   
		(7, 4.58661417e-05)   
		(8, 2.36220472e-06)
	};
	\addlegendentry{BP-FF (Multiloss)}
	
	\end{semilogyaxis}
	\end{tikzpicture}
	\caption{BER results for BCH(127,99) code trained with cycle reduced parity check matrix.}
	\label{fig:bch_127_99_ber_iregular}
\end{figure}
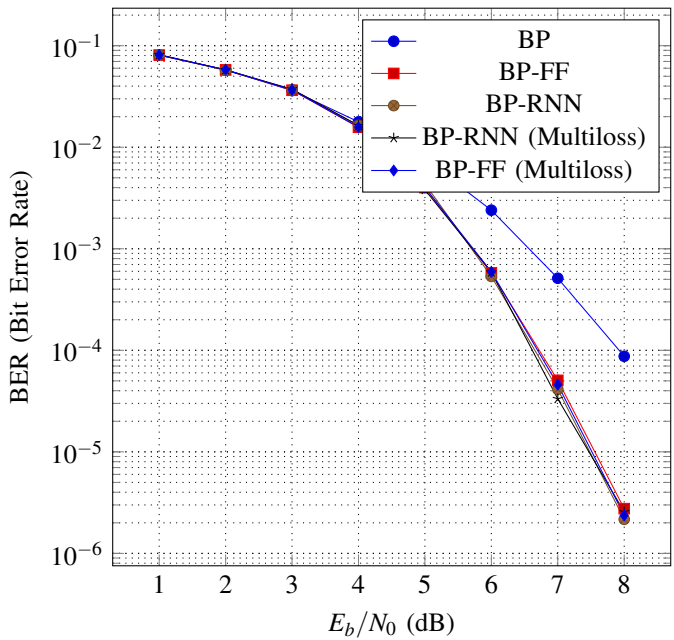

In Figures~\ref{fig:bch_127_64_ber_iregular}, \ref{fig:bch_127_99_ber_iregular} we provide the bit-error-rate for BCH code with $N=127$ for cycle reduced parity check matrix based on \cite{cycle_reduce}. For BCH(127,64) and BCH(127,99) we get an improvement up to $0.9{\rm dB}$ and $1.0{\rm dB}$ respectively.

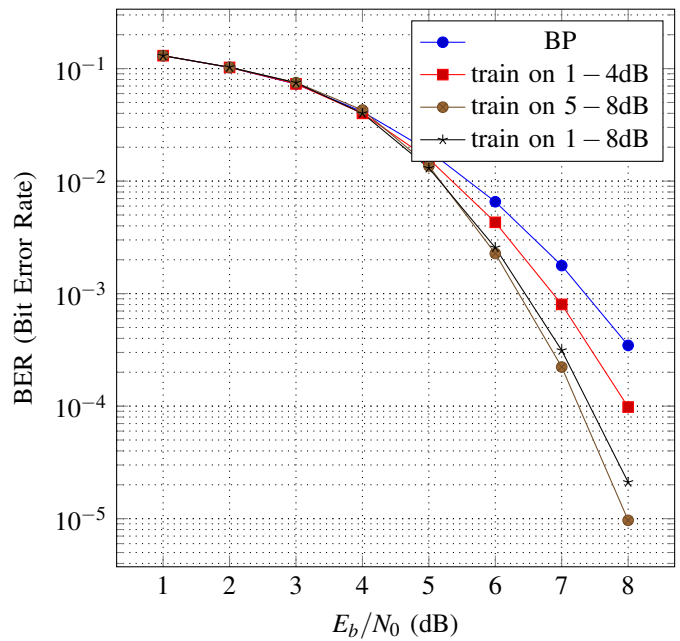
\begin{figure}
	\centering	
	\begin{tikzpicture}
	\begin{semilogyaxis}[
	height=9cm,
	width=9cm,
	grid=both,
	xlabel=$E_b / N_0$ (dB),
	ylabel=BER (Bit Error Rate)
	]
	
	\addplot coordinates {
		(1, 0.13023622)  
		(2, 0.10232283)  
		(3, 0.07292913)  
		(4, 0.04138386)  
		(5, 0.01860433)  
		(6, 0.0065561)
		(7, 0.00177579)  
		(8, 0.00034665)
	};
	\addlegendentry{BP}
	
	\addplot coordinates {
		(1, 1.30263780e-01)   
		(2, 1.02438976e-01)   
		(3, 7.31889764e-02)   
		(4, 3.99960630e-02)
		(5, 1.57440945e-02)   
		(6, 4.30177165e-03)   
		(7, 8.01968504e-04)   
		(8, 9.81692913e-05)
	};
	\addlegendentry{train on $1-4{\rm dB}$}

	\addplot coordinates {
		(1, 1.30289370e-01)   
		(2, 1.02627953e-01)   
		(3, 7.51220472e-02)   
		(4, 4.28346457e-02)
		(5, 1.36397638e-02)   
		(6, 2.27303150e-03)   
		(7, 2.22637795e-04)   
		(8, 9.66535433e-06)
	};
	\addlegendentry{train on $5-8{\rm dB}$}

	\addplot coordinates {
		(1, 1.30100394e-01 )
		(2,   1.02824803e-01 )
		(3,   7.48877953e-02 )
		(4,   4.00452756e-02 )
		(5,   1.30570866e-02 )
		(6,   2.56299213e-03  )
		(7,  3.13188976e-04  )
		(8,  2.10629921e-05)
	};
	\addlegendentry{train on $1-8{\rm dB}$}
		
	\end{semilogyaxis}
	\end{tikzpicture}
	\caption{BER results for BCH(127,64) code trained with cycle reduced parity check matrix. We compare plain BP to BP-RNN with a varying training SNR range.}
	\label{fig:bch_127_64_ber_ireg_SNR_range}
\end{figure}

We assume that the channel is known. Plain BP also assumes a known channel in order to compute the LLRs from the channel observations. In addition to that, in order to train the neural decoder, we are using a varying SNR range for creating noisy codewords which are the input to the training algorithm. To assess the robustness with respect to the training SNR range, we trained the BP-RNN decoder for the BCH(127,64) code with a cycle reduced parity check matrix, using the following three SNR ranges: $1-4{\rm dB}$, $5-8{\rm dB}$ and the full range $1-8{\rm dB}$ (as in Figure \ref{fig:bch_127_64_ber_iregular}). The result is shown in Figure \ref{fig:bch_127_64_ber_ireg_SNR_range}. As can be seen, the performance can be further improved by properly choosing the SNR range in the training, so that it would match the region of interest in actual test conditions. 

\subsection{Neural min-sum decoders}
Next, we present results for decoders which use the min-sum approximation. We trained neural min-sum decoders using the Adam optimizer \cite{ADAM}, with multiloss and a learning rate of 0.1 for the NOMS decoders and 0.01 for the NNMS decoders. All other parameters were set to their default values in the Tensorflow implementation of the Adam optimizer.

\begin{figure}
    \centering
\begin{tikzpicture}
        \begin{semilogyaxis}[
            height=11cm,
            width=9cm,
            ymax=1,
            grid=both,
            xlabel=$E_b / N_0$ (dB),
            ylabel=BER (Bit Error Rate)
        ]

        \addplot coordinates {
            (1,0.08903147323411112)
            (2, 0.0616703462626724)
            (3, 0.03522166393869032)
            (4, 0.01705365866037329)
            (5, 0.0071019374976209505)
            (6, 0.002409659573928159)
            (7, 0.0005251354471977973)
            (8, 0.00010214050981437073)
        };
        \addlegendentry{BP}

        \addplot coordinates {
            (1,0.11787966452235038)
            (2, 0.0911984697955921)
            (3, 0.060940771192569756)
            (4, 0.031682907642140254)
            (5, 0.011872089629883394)
            (6, 0.003248036491441767)
            (7, 0.0007436907616763732)
            (8, 0.00010499536878433762)
        };
        \addlegendentry{Min-sum}
    
        \addplot coordinates {
            (1, 8.95687831e-02)
            (2,   6.18478836e-02)
            (3,   3.37261905e-02)
            (4,   1.26177249e-02)
            (5,3.08994709e-03   )
            (6,4.67328042e-04   )
            (7,4.77513228e-05  )
            (8, 2.91005291e-06)
        };
        \addlegendentry{BP-FF}
        
        \addplot coordinates {
            (1,8.96865079e-02   )
            (2,6.26349206e-02   )
            (3,3.49742063e-02   )
            (4,1.32480159e-02)
            (5, 3.16369048e-03   )
            (6,4.91369048e-04   )
            (7,4.88095238e-05   )
            (8,3.96825397e-06)
        };
        \addlegendentry{BP-RNN}        

        \addplot coordinates {
            (1,0.08979007270374177)
            (2, 0.06482084173930697)
            (3, 0.03918785606435487)
            (4, 0.01689489678098791)
            (5, 0.004397593036681766)
            (6, 0.0006956339690152639)
            (7, 9.56377754938906e-05)
            (8, 1.1773690160707618e-05)
        };
        \addlegendentry{NOMS-FF}
        
        \addplot coordinates {
            (1,0.0901269460622)
            (2,0.0650823785416)
            (3,0.0392925342266)
            (4,0.0172278050575)
            (5,0.00483993757375)
            (6,0.000913396267113)
            (7,0.000146073617297)
            (8,2.45951456094e-05)
        };
        \addlegendentry{NOMS-RNN}
        
        \addplot coordinates {
            (1,0.08894106936672884)
            (2, 0.061739021481227715)
            (3, 0.035813095555301794)
            (4, 0.015085550607133342)
            (5, 0.004064050347023968)
            (6, 0.0006079263573268369)
            (7, 4.8373999213327754e-05)
            (8, 1.905820195505762e-06)
        };
        \addlegendentry{NNMS-FF}
                
        \addplot coordinates {
            (1,0.08965573572887721)
            (2, 0.062278748429827564)
            (3, 0.03603244388616091)
            (4, 0.015036859401367795)
            (5, 0.004010759646251253)
            (6, 0.0006015822262824661)
            (7, 5.0118635250529734e-05)
            (8, 2.3327771992490323e-06)
        };
        \addlegendentry{NNMS-RNN}

        \end{semilogyaxis}
    \end{tikzpicture}
    \caption{Performance comparison of BP and min-sum decoders for BCH (63,45) code.}
    \label{fig:BCH_63_45_minsum}
\end{figure}
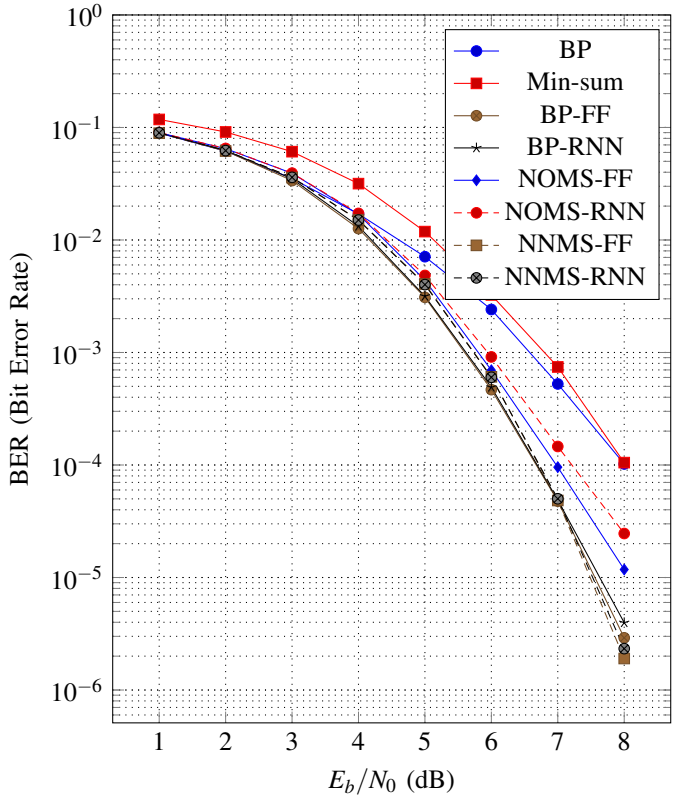

Figure \ref{fig:BCH_63_36_minsum} and Figure \ref{fig:BCH_63_45_minsum} show the BER performance for the BCH(63,36) and BCH(63,45) codes, respectively, with non-sparsified parity check matrices. It can be seen from these plots that the decoders with multiplicative weights (BP-FF, BP-RNN, NNMS-FF, NNMS-RNN) achieve similar performance, implying that the min-sum approximation has little impact. It is also evident that decoders with multiplicative weights outperform decoders with additive offsets (NOMS-FF, NOMS-RNN), although the NOMS decoders still substantially outperform the non-neural decoders.

\begin{figure}
    \centering
\begin{tikzpicture}
        \begin{semilogyaxis}[
            height=11cm,
            width=9cm,
            ymax=1,
            grid=both,
            xlabel=$E_b / N_0$ (dB),
            ylabel=BER (Bit Error Rate)
        ]

        \addplot coordinates {
            (1,0.113)
            (2, 0.0817)
            (3, 0.0491)
            (4, 0.0243)
            (5, 0.0103)
            (6, 0.00344)
            (7, 0.00103)
            (8, 0.0001995)
        };
        \addlegendentry{BP}

        \addplot coordinates {
            (1,0.147)
            (2, 0.11763)
            (3, 0.082048)
            (4, 0.04571)
            (5, 0.019066)
            (6, 0.005983)
            (7, 0.001368)
            (8, 0.00028849)
        };
        \addlegendentry{Min-sum}
    
        \addplot coordinates {
            (1,1.13834656e-01   )
            (2,8.24378307e-02   )
            (3,4.66917989e-02    )
            (4, 1.89285714e-02)
            (5,    5.00529101e-03   )
            (6, 8.21957672e-04   )
            (7, 7.65873016e-05   )
            (8, 5.15873016e-06 )
        };
        \addlegendentry{BP-FF}
        
        \addplot coordinates {
            (1,1.13958333e-01   )
            (2,8.26855159e-02   )
            (3,4.80625000e-02   )
            (4,1.89117063e-02)
            (5, 4.87500000e-03)
            (6,   8.39285714e-04)
            (7,   9.06746032e-05)
            (8,   6.54761905e-06)
        };
        \addlegendentry{BP-RNN}        

        \addplot coordinates {
            (1,0.11491425906893533)
            (2, 0.08705289736464797)
            (3, 0.055978391889662873)
            (4, 0.026183339043051274)
            (5, 0.007501776356692424)
            (6, 0.0011985649575577634)
            (7, 0.00014226713867001637)
            (8, 1.6471576201420018e-05)
        };
        \addlegendentry{NOMS-FF}
        
        \addplot coordinates {
            (1,0.11600576681511933)
            (2, 0.08786843541040183)
            (3, 0.056726840749622524)
            (4, 0.026946696610965198)
            (5, 0.00805752223617931)
            (6, 0.0014334564094755941)
            (7, 0.00017224315785466864)
            (8, 2.3853932726834404e-05)
        };
        \addlegendentry{NOMS-RNN}
        
        \addplot coordinates {
            (1,0.11323702942407979)
            (2, 0.0821558626114981)
            (3, 0.049998889777067235)
            (4, 0.02247281539847487)
            (5, 0.006434693515029247)
            (6, 0.001027114816083641)
            (7, 8.231510030071181e-05)
            (8, 3.3729921195502424e-06)
        };
        \addlegendentry{NNMS-FF}
                
        \addplot coordinates {
            (1,0.11428317663329654)
            (2, 0.08284467663964067)
            (3, 0.05037906182990116)
            (4, 0.022498667732480684)
            (5, 0.0064170885513811175)
            (6, 0.0010321901209191377)
            (7, 8.62801822034436e-05)
            (8, 3.6541048586086554e-06)
        };
        \addlegendentry{NNMS-RNN}

        \end{semilogyaxis}
    \end{tikzpicture}
    \caption{Performance comparison of BP and min-sum decoders for BCH (63,36) code.}
    \label{fig:BCH_63_36_minsum}
\end{figure}
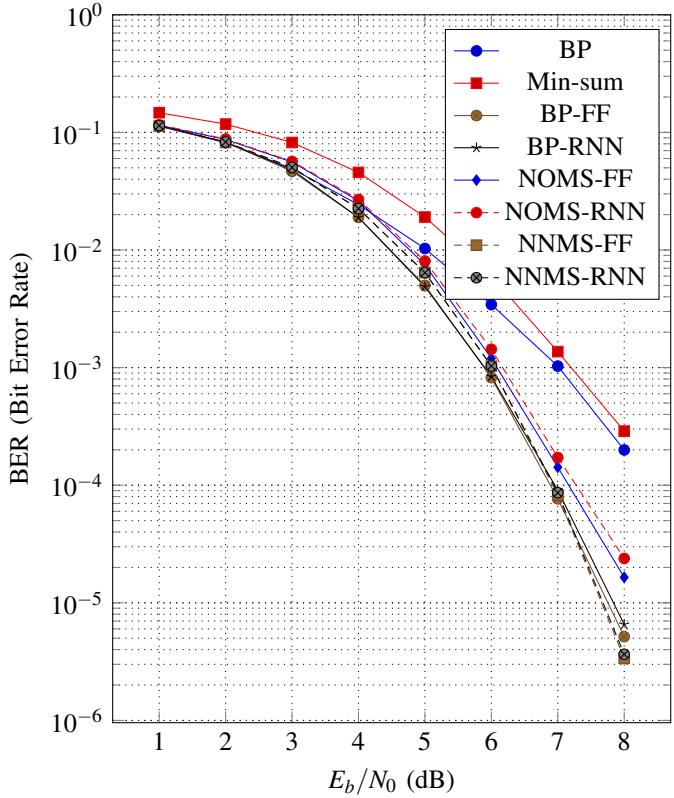

\begin{figure}
    \centering
\begin{tikzpicture}
        \begin{semilogyaxis}[
            height=9cm,
            width=9cm,
            grid=both,
            xlabel=$E_b / N_0$ (dB),
            ylabel=BER (Bit Error Rate)
        ]

        \addplot coordinates {
            (1,0.11787966452235038)
            (2, 0.0911984697955921)
            (3, 0.060940771192569756)
            (4, 0.031682907642140254)
            (5, 0.011872089629883394)
            (6, 0.003248036491441767)
            (7, 0.0007436907616763732)
            (8, 0.00010499536878433762)
        };
        \addlegendentry{Min-sum}
    
        \addplot coordinates {
            (1,0.09255436920305025)
            (2,0.06556025021252838)
            (3,0.03916549300242346)
            (4,0.01711852740030198)
            (5,0.0047757032469262686)
            (6,0.0007579650565262076)
            (7,6.217248423483436e-05)
            (8,2.4390737230854162e-06)
        };
        \addlegendentry{$\gamma = 0.863$}
        
        \addplot coordinates {
            (1,0.09244176087701267)
            (2,0.06561576135916664)
            (3,0.03904162384378212)
            (4,0.01716436374709756)
            (5,0.004790136145052212)
            (6,0.0007722393513760421)
            (7,6.629616941367541e-05)
            (8,2.5405170406442347e-06)
        };
        \addlegendentry{$\gamma = 0.875$}   
        
        \addplot coordinates {
            (1,0.08922052833923337)
            (2, 0.06312061461941558)
            (3, 0.03651110857345869)
            (4, 0.014438925050435841)
            (5, 0.0033457361095250784)
            (6, 0.000432352530673873)
            (7, 3.007648019248947e-05)
            (8, 1.48396749435321e-06)
        };
        \addlegendentry{Relaxed NOMS-FF}  
        
        \end{semilogyaxis}
    \end{tikzpicture}
    \caption{Performance of a min-sum decoder (i.e., $\gamma = 0$), a relaxed min-sum decoder which has learned $\gamma = 0.863$, a relaxed min-sum decoder which is constrained to using $\gamma = 0.875$, and a relaxed NOMS decoder, all for the BCH (63,45) code.}
    \label{fig:BCH_63_45_relaxed}
\end{figure}
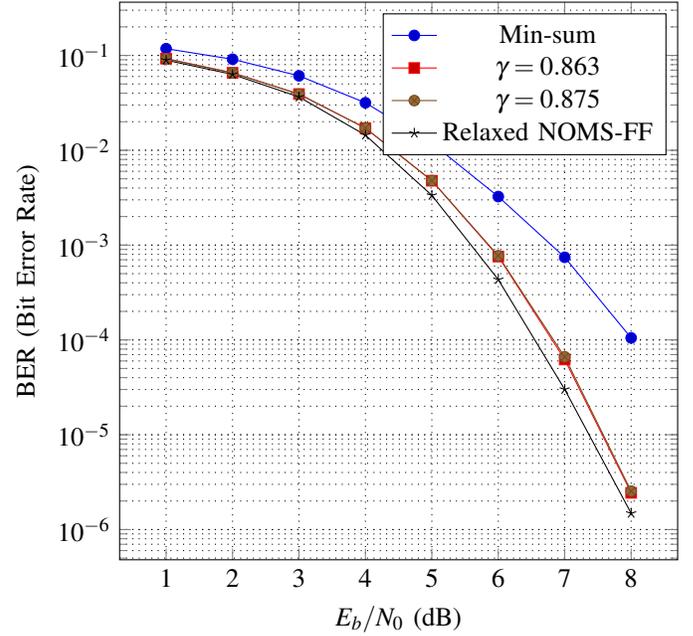

We also present results for an experiment with relaxed decoders. Fig.~\ref{fig:BCH_63_45_relaxed} compares the performance of a simple min-sum decoder with that of three relaxed min-sum decoders trained using the Adam optimizer with a learning rate of 0.01. All relaxed decoders outperform the simple min-sum decoder, achieving a coding gain similar to some of the other decoders presented here.

\begin{figure}
    \centering

    \begin{tikzpicture}
    \begin{axis}[
        width=\linewidth, 
        thick,
        no markers,
        grid=major, 
        grid style={dashed,gray!30},
        xlabel=Number of minibatches processed,
        ylabel=Value of $\gamma$
    ]
    \addplot table [x, y, col sep=comma]{learning_curve.csv}; 
    \end{axis}
    \end{tikzpicture}

    \caption{Evolution of the learnable parameter $\gamma$ as training proceeds.}
    \label{fig:learning_curve}
\end{figure}
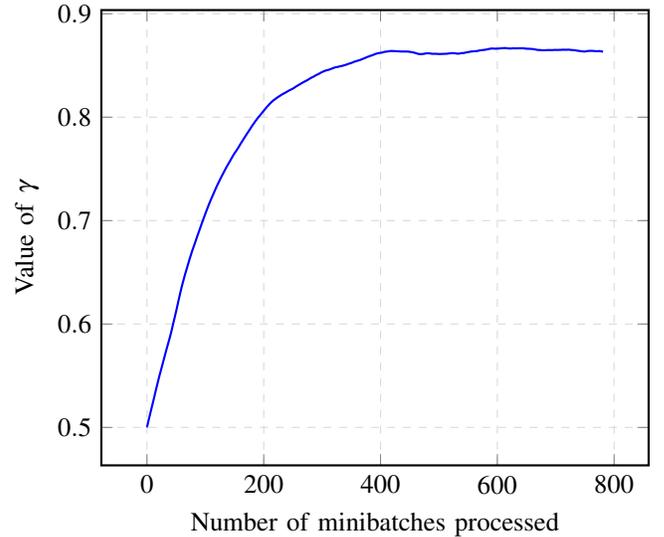

The evolution of the relaxation parameter $\gamma$ as training proceeds is plotted in Fig.~\ref{fig:learning_curve}. It can be seen in this figure that the first relaxed decoder learns a relaxation factor of roughly 0.863. However, note that 0.863 is close to 0.875, and by constraining the second relaxed decoder to using $\gamma = 0.875$ instead of 0.863, we can rewrite \eqref{eq:relax} as follows:
\begin{equation}
    m_t' = m_{t-1}' + 0.125 (m_t - m_{t-1}'),
\end{equation}
which can be implemented very efficiently in hardware, since 0.125 is a power of two ($2^{-3}$) and requires no multiplier. As Fig.~\ref{fig:BCH_63_45_relaxed} shows, constraining the learned relaxation factor in this case causes a nearly imperceptible increase in BER. The third decoder is a relaxed NOMS decoder; it achieves even better performance than the other two relaxed decoders, showing that relaxation can be successfully combined with the learnable decoder building blocks described in \cite{nachmani} and \cite{lugosch}.

Fig.~\ref{fig:learning_curve} shows that the parameter $\gamma$ attains its final value after 400 minibatches of 120 frames each have been processed. Since the number of operations performed during the forward pass of training (belief propagation) is roughly equal to the number of operations performed during the backward pass of training (backpropagation), training the decoder in this case requires processing the equivalent of $400 \times 120 \times 2 = $ 96,000 frames. In contrast, simulating the decoder until 100 frame errors have occurred at an SNR of $8{\rm dB}$ requires processing 1,064,040 frames. Naturally, testing the decoder with multiple different candidate values for $\gamma$ in search of the optimal value will require processing even more frames.
In deep learning, gradient descent is usually applied to neural networks with a very large number of parameters. Our results for the relaxed min-sum decoder show that
gradient-based learning can be a very efficient way of optimizing decoder parameters even when the number of parameters is small.

\subsection{mRRD-RNN and mRRD-NOMS} 
Finally, we provide the bit error rate results when using our proposed mRRD-RNN (BP version) decoder and when using a similar mRRD-NOMS decoder applied to a BCH(63,36) code represented by a cycle reduced parity check matrix based on \cite{cycle_reduce}. In the experiments we use the BP-RNN with multiloss architecture and an unfold of 5, which corresponds to 5 BP iterations. The parameters of the mRRD-RNN are as follows. We use 2 iterations for each ${\rm BP}_{i,j}$ block in Figure~\ref{fig:mrrd_diag}, a value of $m=1,3,5,50$, denoted in the following by mRRD-RNN($m$), and a value of $c=30$ ($c=50$, respectively) when $m=1,3,5$ ($m=50$). We also experimented with a similar mRRD-NOMS($m$) algorithm, using the NOMS-RNN (Equation \eqref{eq:x_ie_LB_NOMS_RNN}), and the same setting of parameters as in the mRRD-RNN algorithm.

\begin{figure}[thpb]
	\centering
		\begin{tikzpicture}
        \begin{semilogyaxis}[
            height=12cm,
            width=9cm,
            ymax=10,
            grid=both,
            xlabel=$E_b / N_0$ (dB),
            ylabel=BER (Bit Error Rate)
        ]
        \addplot[mark=star,mark options={fill=violet},color=violet] coordinates {
            (3, 0.0258 )
            (3.5, 0.0144)
            (4,  0.0068)
            (4.5,  0.0029)
            (5, 9.1278e-04)
        };
        \addlegendentry{HDD}
        
        \addplot[mark=triangle*,mark options={fill=red},color=red] coordinates {
			(3, 0.0248638)
			(3.5, 0.0108071)
			(4, 0.00461746)
			(4.5, 0.00141587)
			(5, 0.000418783)
        };
        \addlegendentry{mRRD(1)}

        \addplot[mark=triangle*,mark options={fill=red},color=red, dashed] coordinates {
			(3, 0.0119249)
			(3.5, 0.00472937)
			(4, 0.00155899)
			(4.5, 0.000425397)
			(5, 9.78307e-05)
		};
        \addlegendentry{mRRD-RNN(1)}
        
        \addplot[mark=*,mark options={fill=green},color=green] coordinates {
			(3, 0.00825608)
			(3.5, 0.00303333)
			(4, 0.000888624)
			(4.5, 0.000186508)
			(5, 4.07407e-05)
        };
        \addlegendentry{mRRD(3)}

        \addplot[mark=*,mark options={fill=green}, color=green, dashed] coordinates {
			(3, 0.00516603)
			(3.5, 0.00167156)
			(4, 0.000423122)
			(4.5, 7.53439e-05)
			(5, 1.34921e-05)
        };
        \addlegendentry{mRRD-RNN(3)}
        
        \addplot[mark=square*,mark options={fill=blue},color=blue] coordinates {
			(3, 0.00586693)
			(3.5, 0.00190397)
			(4, 0.00050291)
			(4.5, 0.000105291)
			(5, 1.85185e-05)
        };
        \addlegendentry{mRRD(5)}

        \addplot[mark=square*,mark options={fill=blue},color=blue, dashed] coordinates {
			(3, 0.00401103)
			(3.5, 0.00120333)
			(4, 0.000261296)
			(4.5, 4.28042e-05)
			(5, 4.65608e-06)			
        };
        \addlegendentry{mRRD-RNN(5)}

		\addplot[mark=diamond*,mark options={fill=brown},color=brown] coordinates {
			(3, 0.00159841)
			(3.5, 0.000358764)
			(4, 6.10871e-05)
			(4.5, 8.89851e-06)
		};
		\addlegendentry{mRRD(50)}

		\addplot[mark=diamond*,mark options={fill=brown},color=brown, dashed] coordinates {
			(3, 0.00140426)
			(3.5, 0.000331758)
			(4, 4.78307e-05)
			(4.5, 5.63492e-06)
		};
		\addlegendentry{mRRD-RNN(50)}

        \addplot[mark=*,mark options={fill=black}, color=black] coordinates {
            (3, 0.000971000000000000)
            (3.5, 0.000222000000000000)
            (4, 3.15000000000000e-05)
            (4.5, 4.21000000000000e-06)
            (5, 3.31000000000000e-07)
        };
        \addlegendentry{ML}
        
        \end{semilogyaxis}
    \end{tikzpicture}
	\caption{mRRD-RNN BER results for BCH(63,36) code.}
	\label{fig:mrrd_ber}
\end{figure}
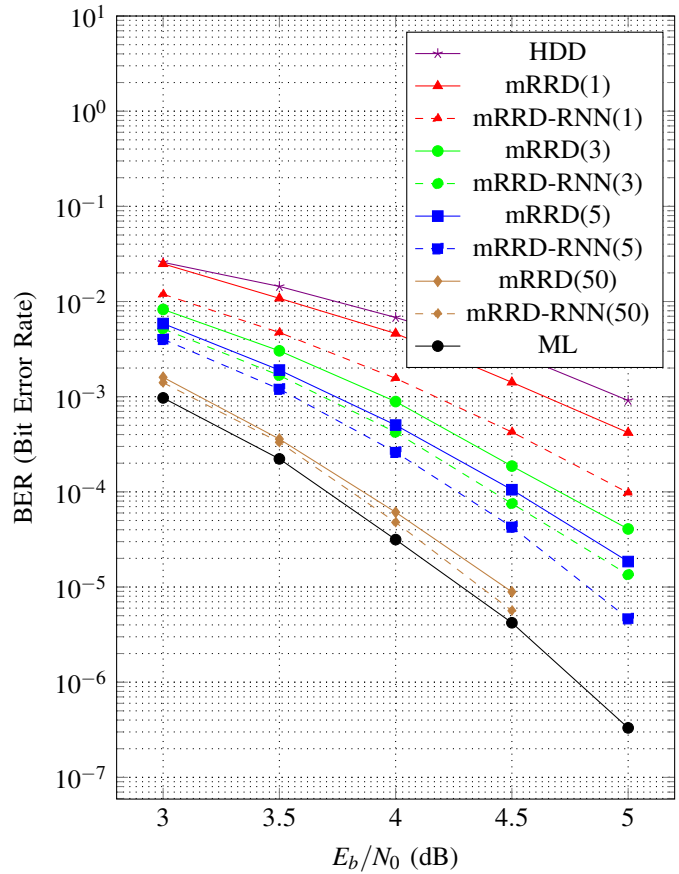

In Figure~\ref{fig:mrrd_ber} we present the bit error rates for mRRD-RNN(1), mRRD-RNN(3), mRRD-RNN(5) and mRRD-RNN(50), and compare it to hard decision decoding (HDD) and to plain mRRD with the same parameters. As can be seen, we achieve improvements of $0.6$dB, $0.3$dB and $0.2$dB compared to plain mRRD for $m=1,2,3$. Hence, the mRRD-RNN decoder can improve on the plain mRRD decoder. The performance of the ML decoder was estimated using the implementation of \cite{Boutros} based on the ordered statistics decoder (OSD) algorithm \cite{fossOSD} (see also \cite{liva2016code}). Note that by increasing the value of $m$, the gap to ML performance decreases towards zero.

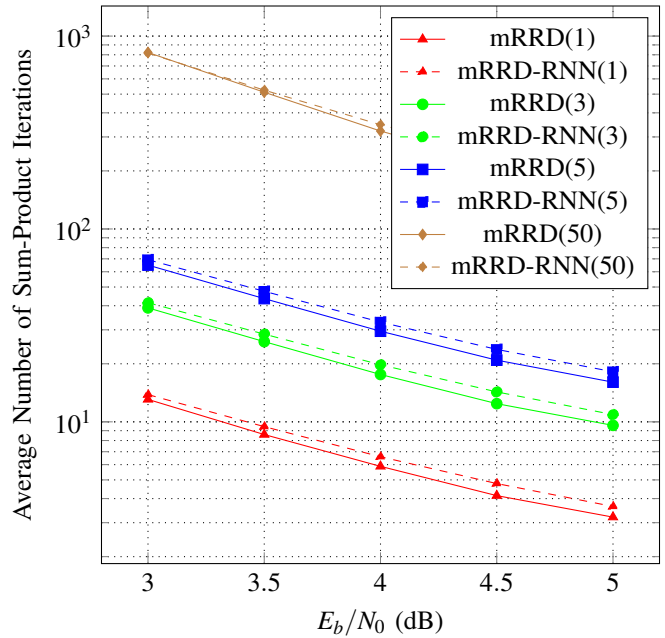
\begin{figure}[thpb]
	\centering
	\begin{tikzpicture}
        \begin{semilogyaxis}[
            height=9cm,
            width=9cm,
            grid=both,
            xlabel=$E_b / N_0$ (dB),
            ylabel=Average Number of Sum-Product Iterations
        ]
        \addplot[mark=triangle*,mark options={fill=red},color=red] coordinates {
			(3, 13.0718)
			(3.5, 8.59376)
			(4, 5.87809)
			(4.5, 4.14426)
			(5, 3.20625)
        };
        \addlegendentry{mRRD(1)}

        \addplot[mark=triangle*,mark options={fill=red},color=red,dashed] coordinates {
            (3, 13.8455    )
            (3.5, 9.44395   )
            (4, 6.59196    )
            (4.5, 4.7868  )
            (5, 3.64153)
        };
        \addlegendentry{mRRD-RNN(1)}
        
        \addplot[mark=*,mark options={fill=green},color=green] coordinates {
            (3, 38.9828 )
            (3.5, 26.0517 )
            (4, 17.5981 )
            (4.5, 12.4233 )
            (5, 9.586)
        };
        \addlegendentry{mRRD(3)}

        \addplot[mark=*,mark options={fill=green},color=green,dashed] coordinates {
            (3, 41.4602   )
            (3.5, 28.4947  )
            (4, 19.7628 )
            (4.5, 14.2974 )
            (5, 10.9036)
        };
        \addlegendentry{mRRD-RNN(3)}
        
        \addplot[mark=square*,mark options={fill=blue},color=blue] coordinates {
            (3, 64.9662 )
            (3.5, 43.5406)
            (4, 29.5024 )
            (4.5, 20.8842)
            (5, 16.06)
        };
        \addlegendentry{mRRD(5)}

        \addplot[mark=square*,mark options={fill=blue},color=blue, dashed] coordinates {
            (3, 69.27    )
            (3.5, 47.5175 )
            (4, 32.8972 )
            (4.5, 23.7465 )
            (5, 18.1812)
        };
        \addlegendentry{mRRD-RNN(5)}
        
        \addplot[mark=diamond*,mark options={fill=brown},color=brown] coordinates {
        	(3, 822.3)
        	(3.5, 511.469)
        	(4, 321.917)
        	(4.5, 218.26)
        };
        \addlegendentry{mRRD(50)}
        
        \addplot[mark=diamond*,mark options={fill=brown},color=brown, dashed] coordinates {
        	(3, 815.512)
        	(3.5, 524.492)
        	(4, 346.869)
        	(4.5, 243.625)
        };
        \addlegendentry{mRRD-RNN(50)}
        
        \end{semilogyaxis}
    \end{tikzpicture}
	\caption{mRRD-RNN complexity results for BCH(63,36).}
	\label{fig:mrrd_comp}
\end{figure}

Figure~\ref{fig:mrrd_comp} compares the average number of BP iterations for the various decoders using plain mRRD and mRRD-RNN. As can be seen, there is a small increase in the complexity of up to 8\% when using the RNN decoder. However, overall, with the RNN decoder one can achieve the same error rate with a significantly smaller computational complexity due to the reduction in the required value of $m$. This improvement decreases when $m$ increases.

\begin{figure}[thpb]
	\centering
	\begin{tikzpicture}
	\begin{semilogyaxis}[
	height=11cm,
	width=9cm,
	ymax=500000,
	grid=both,
	xlabel=$E_b / N_0$ (dB),
	ylabel=Decoding Time ($\mu$-sec)
	]

	\addplot[mark=triangle*,mark options={fill=red},color=red] coordinates {
		(3, 466.536)
		(3.5, 306.412)
		(4, 208.953)
		(4.5, 141.86)
		(5, 98.9969)
	};
	\addlegendentry{mRRD(1)}
	
	\addplot[mark=triangle*,mark options={fill=red},color=red,dashed] coordinates {
		(3, 281.007)
		(3.5, 187.874)
		(4, 136.1)
		(4.5, 86.1848)
		(5, 63.0983)
	};
	\addlegendentry{mRRD-NOMS(1)}
	
	\addplot[mark=*,mark options={fill=green},color=green] coordinates {
		(3, 1398.19)
		(3.5, 916.717)
		(4, 611.659)
		(4.5, 415.27)
		(5, 299.764)
	};
	\addlegendentry{mRRD(3)}
	
	\addplot[mark=*,mark options={fill=green},color=green,dashed] coordinates {
		(3, 785.507)
		(3.5, 536.894)
		(4, 355.488)
		(4.5, 245.29)
		(5, 175.416)
	};
	\addlegendentry{mRRD-NOMS(3)}
	
	\addplot[mark=square*,mark options={fill=blue},color=blue] coordinates {
		(3, 2324.97)
		(3.5, 1530.02)
		(4, 1019.34)
		(4.5, 696.869)
		(5, 509.198)
	};
	\addlegendentry{mRRD(5)}
	
	\addplot[mark=square*,mark options={fill=blue},color=blue, dashed] coordinates {
		(3, 1485.08)
		(3.5, 964.154)
		(4, 633.134)
		(4.5, 399.585)
		(5, 285.567)
	};
	\addlegendentry{mRRD-NOMS(5)}

	\addplot[mark=diamond*,mark options={fill=brown},color=brown] coordinates {
		(3, 31661)
		(3.5, 18315.6)
		(4, 11023.5)
		(4.5, 7072.17)
	};
	\addlegendentry{mRRD(50)}

	\addplot[mark=diamond*,mark options={fill=brown},color=brown, dashed] coordinates {
		(3, 15635)
		(3.5, 9746.58)
		(4, 6085.85)
		(4.5, 3939.48)
	};
	\addlegendentry{mRRD-NOMS(50)}
	
	\end{semilogyaxis}
	\end{tikzpicture}
	\caption{mRRD-NOMS decoding time for BCH(63,36).}
	\label{fig:mrrd_noms_time}
\end{figure}
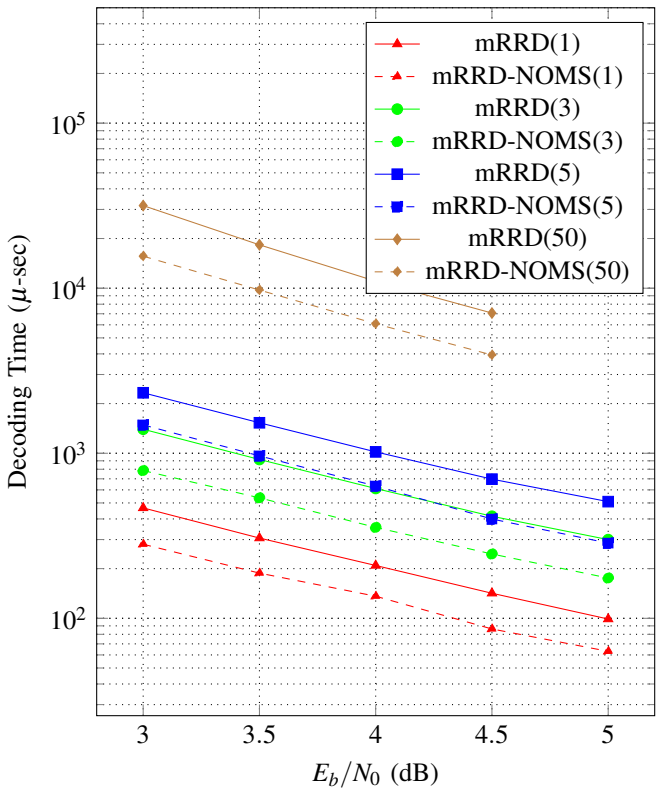

Figures \ref{fig:mrrd_noms_ber} and \ref{fig:mrrd_noms_time} present a similar comparison between the mRRD and the mRRD-NOMS decoders. The NOMS decoder used is described by a modified version of Equation \eqref{eq:x_ie_LB_NOMS_RNN}, which is now multiplied by a fixed weight of $1/2$ as in \eqref{eq:x_ie_LB_NMS}. The motivation for using this fixed attenuation is the same as in the NMS algorithm (see the motivating argument for \eqref{eq:x_ie_LB_NMS} above). As can be seen, the mRRD-NOMS decoder improves the corresponding mRRD decoder with the same parameters both with respect to error rate and with respect to decoding time, throughout the SNR region that was examined.

\begin{figure}[thpb]
	\centering	
	\begin{tikzpicture}
	\begin{semilogyaxis}[
	height=11cm,
	width=9cm,
	ymax=15,
	grid=both,
	xlabel=$E_b / N_0$ (dB),
	ylabel=BER (Bit Error Rate)
	]
	
	\addplot[mark=triangle*,mark options={fill=red},color=red] coordinates {
		(3, 0.0248638)
		(3.5, 0.0108071)
		(4, 0.00461746)
		(4.5, 0.00141587)
		(5, 0.000418783)
	};
	\addlegendentry{mRRD(1)}
	
	\addplot[mark=triangle*,mark options={fill=red},color=red, dashed] coordinates {
		(3, 0.0128304)
		(3.5, 0.00544423)
		(4, 0.00190352)
		(4.5, 0.000549444)
		(5, 0.000129815)
	};
	\addlegendentry{mRRD-NOMS(1)}
	
	\addplot[mark=*,mark options={fill=green},color=green] coordinates {
		(3, 0.00825608)
		(3.5, 0.00303333)
		(4, 0.000888624)
		(4.5, 0.000186508)
		(5, 4.07407e-05)
	};
	\addlegendentry{mRRD(3)}
	
	\addplot[mark=*,mark options={fill=green}, color=green, dashed] coordinates {
		(3, 0.00569233)
		(3.5, 0.00194101)
		(4, 0.000543386)
		(4.5, 0.000111481)
		(5, 2e-05)
	};
	\addlegendentry{mRRD-NOMS(3)}
	
	\addplot[mark=square*,mark options={fill=blue},color=blue] coordinates {
		(3, 0.00586693)
		(3.5, 0.00190397)
		(4, 0.00050291)
		(4.5, 0.000105291)
		(5, 1.85185e-05)
	};
	\addlegendentry{mRRD(5)}
	
	\addplot[mark=square*,mark options={fill=blue},color=blue, dashed] coordinates {
		(3, 0.00463333)
		(3.5, 0.00142354)
		(4, 0.000282804)
		(4.5, 5.65079e-05)
		(5, 9.20635e-06)
	};
	\addlegendentry{mRRD-NOMS(5)}

	\addplot[mark=diamond*,mark options={fill=brown},color=brown] coordinates {
		(3, 0.00159841)
		(3.5, 0.000358764)
		(4, 6.10871e-05)
		(4.5, 8.89851e-06)
	};
	\addlegendentry{mRRD(50)}

	\addplot[mark=diamond*,mark options={fill=brown},color=brown, dashed] coordinates {
		(3, 0.00156878)
		(3.5, 0.000325875)
		(4, 5.48822e-05)
		(4.5, 6.34921e-06)
	};
	\addlegendentry{mRRD-NOMS(50)}
	
	\addplot[mark=*,mark options={fill=black}, color=black] coordinates {
		(3, 0.000971000000000000)
		(3.5, 0.000222000000000000)
		(4, 3.15000000000000e-05)
		(4.5, 4.21000000000000e-06)
		(5, 3.31000000000000e-07)
	};
	\addlegendentry{ML}
	
	\end{semilogyaxis}
	\end{tikzpicture}
	\caption{mRRD-NOMS BER results for BCH(63,36) code.}
	\label{fig:mrrd_noms_ber}
\end{figure}
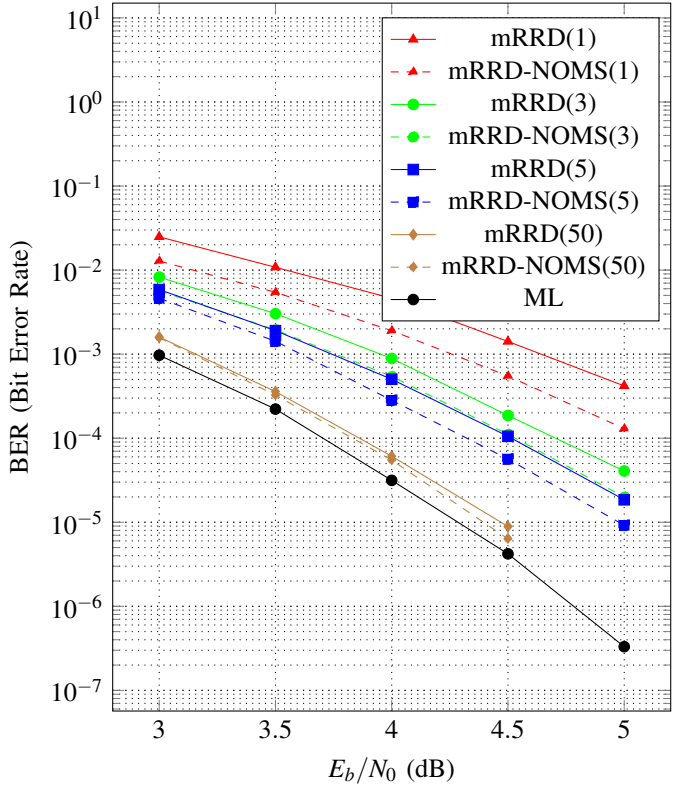 

\section{Complexity and Comparison with other methods}
When the block length is sufficiently large, LDPC codes under BP decoding can be used for efficient reliable communication, very close to channel capacity. However, as the block length decreases, other approaches can perform better.
In particular, when transmitting over the BIAWGNC, BCH codes are very close to the best possible error rate for the given channel, block length and rate~\cite{Boutros,liva2016code}.

In order to decode a BCH or any other linear code, one can use the OSD algorithm \cite{fossOSD}. This algorithm requires a Gaussian elimination stage, followed by an exhaustive search over $\sum_{i=0}^d {K \choose i}$ possible codewords whose distances to the received channel observation vector need to be computed. Here, $K$ is the number of information bits and $d$ is a parameter of the algorithm. The OSD algorithm is computationally efficient for some channels, such as the BIAWGNC, and less efficient for other channels, such as the binary symmetric channel (BSC), which is important for applications such as coding for memories. This is due to the fact that computationally efficient decoding using OSD requires soft channel information. Another difficulty with the OSD is that the Gaussian elimination is difficult to implement efficiently in hardware for low latency communications since it is an inherently serial algorithm. 

The RRD~\cite{cycle_reduce}, MBBP~\cite{mbbp} and mRRD~\cite{dimnik2009improved} algorithms have been suggested as alternative low complexity, close to ML algorithms for HDPC codes such as BCH codes. In this work we have shown improvements compared to the mRRD algorithm. Both plain mRRD, and our neural mRRD decoders can be easily implemented in hardware, since the basic operation that needs to be performed is either (neural) BP or (neural) min-sum decoding. Consider the NOMS decoder with parameter tying described by Equations~\eqref{eq:x_ie_RB}, \eqref{eq:x_ie_LB_NOMS} and
and \eqref{eq:ov}. As is well known, in order to implement~\eqref{eq:x_ie_RB} efficiently, one first computes
$$
s_v = l_v + \sum_{e'=(v,c')} x_{i-1,e'}
$$
for each variable node, $v$, in the Tanner graph. Then, for each edge $e=(v,c)$, $x_{i,e}$ is obtained using
$$
x_{i,e} = s_v - x_{i-1,e}
$$
Thus, each iteration, efficient computation of~\eqref{eq:x_ie_RB}  requires about $E$ summations.
A similar idea can be applied in order to implement \eqref{eq:x_ie_LB_NOMS} efficiently, so that $O(E)$ operations are required each iteration (with no multiplications).

Note that in the definition of the BP-RNN decoder, Equation \eqref{eq:x_ie_RB_NN_rnn}, we used the weights $w_{e,e'}$, while in the definitions of the neural min-sum decoders, \eqref{eq:x_ie_LB_NNMS_RNN} and \eqref{eq:x_ie_LB_NOMS_RNN}, we used $w_e$ and $\beta_e$. In fact, it is possible to use $w_{e'}$ instead of $w_{e,e'}$ also in \eqref{eq:x_ie_RB_NN_rnn}, and our experience shows that the error rate obtained is about the same. In this case, the same efficient computation method indicated above for the NOMS decoder can also be used for the BP-RNN decoder, and the computational complexity remains $O(E)$. However, using the NOMS decoder is advantageous since it does not require multiplications.

Finally, we note that in order to scale our decoders to longer block length codes, one can use the polar-concatenated scheme in \cite{mahdavifar2014performance,trifonov2011generalized,wang2016interleaved,goldin2017performance}.
According to this approach, one can construct powerful longer block length codes from shorter constituent codes (e.g., BCH codes with block length 64), and decode them efficiently using computationally efficient close to ML decoders for the shorter constituent codes.
A similar scaling approach for the decoding of polar codes was used in \cite{cammerer2017scaling}.  

\section{Conclusion} 
We introduced neural architectures for decoding linear block codes. These architectures yield significant improvements over the standard BP and min-sum decoders. Furthermore, we showed that the neural network decoder improves on standard BP even for cycle reduced parity check matrices, with improvements of up to $1.5{\rm dB}$ in the SNR. We also showed performance improvement of the mRRD algorithm with the new RNN architecture. We regard this work as a first step towards the design of deep neural network-based decoding algorithms.

The decoders we introduce in this work offer a trade-off between error-correction performance and implementation complexity. For instance, while adders are more hardware-friendly than multipliers, decoders with additive offsets may not perform quite as well as those with multiplicative weights. Relaxed decoders outperform non-relaxed decoders, but relaxation requires additional memory to store a previous iteration's messages. Which decoder one chooses depends on the needs of the application. In every instance, the use of machine learning improves the performance of the decoder.

Our future work includes possible improvements in the performance by exploring new neural network architectures. Moreover, we will investigate end-to-end learning of the mRRD algorithm (i.e. learning graph with permutation), and fine tune the parameters of the mRRD-RNN algorithm.
Another direction is the consideration of an RNN architecture with quantized weights in order to reduce the number of free parameters. It has been shown in the past \cite{Soudry,xnor_net} that in various applications the loss in performance incurred by weight quantization can be small if this quantization is performed properly.
Finally, in some communication systems, e.g. \cite{farsad2017detection}, it is not possible to accurately model the channel. We believe that our proposed methods may be useful in these scenarios. It would also be interesting to consider the case where the channel used in training may deviate from the actual channel in test conditions.





\section*{ACKNOWLEDGMENT}

We thank Jacob Goldberger for his comments on our work, Johannes Van Wonterghem and Joseph J. Boutros for making their OSD software available to us. We also thank Ilan Dimnik, Ofir Liba, Ethan Shiloh and Ilia Shulman for helpful discussion and their support to this research, and Gianluigi Liva for sharing with us his OSD Matlab package.

This research was supported by the Israel Science Foundation, grant no. 1082/13. The Tesla K40c used for this research was donated by the NVIDIA Corporation.

\end{document}